\documentclass{sig-alternate}

\newcommand{\ignore}[1]{}
\usepackage{fancyhdr}
\usepackage[hyphens]{url}
\usepackage[normalem]{ulem}
\usepackage[nocompress]{cite}
\usepackage{epsfig}
\usepackage{graphicx}
\usepackage{subcaption}
\usepackage{caption}
\usepackage{epstopdf}
\usepackage{ragged2e}
\usepackage{hyperref}
\usepackage{amsmath}
\usepackage{algorithm}
\usepackage{algpseudocode}
\usepackage{listings}
\usepackage{glossaries}
\usepackage{easylist}
\usepackage{bbding}
\usepackage[mediumspace,mediumqspace,Grey,squaren]{SIunits}
\usepackage{ctable}
\usepackage[inline]{enumitem}
\setitemize{noitemsep,topsep=0pt,parsep=0pt,partopsep=0pt}
\setenumerate{noitemsep,topsep=0pt,parsep=0pt,partopsep=0pt}
\usepackage[font=small,skip=0pt]{caption}
\usepackage{amssymb}
\usepackage{graphicx}
\usepackage{auto-pst-pdf}

\title{CG-OoO\\Energy-Efficient Coarse-Grain Out-of-Order Execution} 
\author{
Milad~Mohammadi\textsuperscript{$\star$},
Tor~M.~Aamodt\textsuperscript{$\dagger$},
William~J.~Dally\textsuperscript{$\star$$\ddagger$}\\
\textsuperscript{$\star$}Stanford University, 
\textsuperscript{$\dagger$}University of British Columbia, 
\textsuperscript{$\ddagger$}NVIDIA Research\\
\texttt{milad@cs.stanford.edu},
\texttt{aamodt@ece.ubc.ca},
\texttt{dally@stanford.edu}
}

\begin{document}
\maketitle
\setlength{\emergencystretch}{4em}


\begin{abstract}
We introduce the Coarse-Grain Out-of-Order (CG-OoO) general purpose processor
designed to achieve close to In-Order processor energy while maintaining
Out-of-Order (OoO) performance. CG-OoO is an energy-performance proportional
general purpose architecture that scales according to the program
load\footnote{Not to be confused with energy-proportional
designs~\cite{barroso2007case}. Energy-performance proportional scaling refers
to linear change in energy as the processor configuration allows higher peak
performance (Figure~\ref{fig:ep_trend}).}.  Block-level code processing is at
the heart of the this architecture; CG-OoO speculates, fetches, schedules, and
commits code at block-level granularity. It eliminates unnecessary accesses to
energy consuming tables, and turns large tables into smaller and distributed
tables that are cheaper to access. CG-OoO leverages compiler-level code
optimizations to deliver efficient static code, and exploits dynamic
instruction-level parallelism and block-level parallelism.  CG-OoO introduces
Skipahead issue, a complexity effective, limited out-of-order instruction
scheduling model. Through the energy efficiency techniques applied to the
compiler and processor pipeline stages, CG-OoO closes 64\% of the average energy
gap between the In-Order and Out-of-Order baseline processors at the performance
of the OoO baseline. This makes CG-OoO $1.9\times$ more efficient than the OoO
on the energy-delay product inverse metric.
\end{abstract}

\section{Introduction\label{sec:intro}}
This paper revisits the Out-of-Order (OoO) execution model and devises an
alternative model that achieves the performance of the OoO at over 50\% lower
energy cost. Czechowski et al.~\cite{czechowski2014improving} discusses the
energy efficiency techniques used in the recent generations of the Intel CPU
architectures (e.g.  Core i7, Haswell) including Micro-op cache, Loop cache, and
Single Instruction Multiple Data (SIMD) instruction set architecture (ISA).
This paper questions the inherent energy efficiency attributes of the
OoO execution model and provides a solution that is over 50\% more energy
efficient than the baseline OoO. The energy efficiency techniques discussed
in~\cite{czechowski2014improving} can also be applied to the CG-OoO model to
make it even more energy efficient.

Despite the significant achievements in improving energy and performance
properties of the OoO processor in the recent
years~\cite{czechowski2014improving}, studies show the energy and performance
attributes of the OoO execution model remain superlinearly
proportional~\cite{widget,azizi2010energy}. Studies indicate control speculation
and dynamic scheduling technique amount to 88\% and 10\% of the OoO superior
performance compared to the In-Order (InO) processor~\cite{dyn_specul}.
Scheduling and speculation in OoO is performed at instruction granularity
regardless of the instruction type even though they are mainly effective during
unpredictable dynamic events (e.g.  unpredictable cache
misses)~\cite{dyn_specul}. Furthermore, our studies show speculation and dynamic
scheduling amount to 67\% and 51\% of the OoO excess energy compared to the InO
processor. These observations suggest any general purpose processor architecture
that aims to maintain the superior performance of OoO while closing the energy
efficiency gap between InO and OoO ought to implement architectural solutions in
which low energy program speculation and dynamic scheduling are central. 

Our study provide four high level observations. 
\begin{enumerate*}[label=]
    \item First, OoO excess energy is well distributed across all pipeline
stages. Thus, an energy efficient architecture should reduce energy of each
stage.
    \item Second, OoO execution model imposes tight functional dependencies
between stages requiring a solution to enable energy efficiency across all
stages.
    \item Third, as mentioned by others, complexity effective
micro-architectures such as ILDP~\cite{ildp} and Palachara, et
al.~\cite{complexity} enable simpler hardware, such as local and global register
files that improve energy efficiency. A block-level execution model, like
CG-OoO, enables energy efficiency by simplifying complex, energy consuming
modules throughout the pipeline stages.
    \item Fourth, since dynamic scheduling and speculation techniques mainly
benefit unpredictable dynamic events, they should be applied to instructions
selectively. Unpredictable events are hard to detect and design for; however, we
show a hierarchy of scheduling techniques can adjust the processing energy
according to the program runtime state.
\end{enumerate*}

CG-OoO contributes a \textit{hierarchy of scheduling techniques} centered around
clustering instructions; static instruction scheduling organizes instructions at
basic-block level granularity to reduce stalls. The CG-OoO dynamic block
scheduler dispatches multiple code blocks concurrently. Blocks issue
instructions in-order when possible. In case of an unpredictable stall, each
block allows limited out-of-order instruction issue using a complexity effective
structure named \textit{Skipahead}; Skipahead accomplishes this by performing
dynamic dependency checking between a very small collection of instructions at
the head of each code block. Section~\ref{sec:issue_sch} discusses the Skipahead
micro-architecture.

CG-OoO contributes a complexity effective \textit{block-level control speculation}
model that saves speculation energy throughout the entire pipeline by allowing
block-level control speculation, fetch, register renaming bypass, dispatch, and
commit.  Several \textit{front-end} architectures have shown block-level
speculation can be done with high accuracy and low energy
cost~\cite{bliss,ftb,bsa}.

CG-OoO uses a distributed \textit{register file hierarchy} to allow static
allocation of block-level, short-living registers, and dynamic allocation of
long-living registers.

The rest of this paper is organized as follows. Section~\ref{sec:related_work}
presents the related work, Section~\ref{sec:exe_model} describes the CG-OoO
execution model, Section~\ref{sec:arch} discusses the processor architecture,
Section~\ref{sec:method} presents the evaluation methodology,
Section~\ref{sec:eval} provides the evaluation results, and
Section~\ref{sec:conclusion} concludes the paper.


\section{Overview \& Related Work}\label{sec:related_work}
CG-OoO aims to design an energy efficient, high-performance, single-threaded,
processor through targeting a design point where the complexity is nearly as
simple as an in-order and instruction-level parallelism (ILP) is paramount.
Table~\ref{tab:rel_work1} compares several high-level design features that
distinguish the CG-OoO processor from the previous literature.  Unlike others,
CG-OoO's primary objective is energy efficient computing (column 3), thereby
designing several complexity effective (col. 4), energy-aware techniques
including: an efficient register file hierarchy (col. 9), a block-level control
speculation, and a static and dynamic block-level instruction scheduler (col. 7,
8) coupled with a complexity effective out-of-order issue model named
\textit{Skipahead}. CG-OoO is a distributed instruction-queue model (col. 2)
that clusters execution units with instruction queues to achieve an
energy-performance proportional solution (col. 6). 

Braid~\cite{braid} clusters static instructions at sub-basic-block granularity.
Each braid runs in-order as a block of code. Clustering static instructions at
this granularity requires additional control instructions to guarantee program
execution correctness.  Injecting instructions increases instruction cache
pressure and processor energy overhead.  Braid performs instruction-level,
branch prediction, issue and commit.  WiDGET~\cite{widget} is a
power-proportional grid execution design consisting of a decoupled thread
context management and a large set of simple execution units. WiDGET performs
instruction-level dynamic data dependency detection to schedule instructions. In
contract to these proposals, the CG-OoO clusters basic-block instructions
statically such that, at runtime, control speculation, fetch, commit, and squash
are done at block granularity. Furthermore, CG-OoO leverages energy
efficient, limited out-of-order scheduling from each code block (col. 8).

\begin{table}[h]
    \begin{center}
    \small
    \begin{tabular}{| c | l | l | l | l | l | l | l | l | l | l |}
    \hline
    DESIGN & \rotatebox{90}{Distributed {\micro}-architecture/Coarse-Grain}
           & \rotatebox{90}{Energy Modeling}
           & \rotatebox{90}{Complexity Effective Design}
           & \rotatebox{90}{Profiling NOT done}
           & \rotatebox{90}{Pipeline Clustering}
           & \rotatebox{90}{Static \& Dynamic Scheduling Hybrid}
           & \rotatebox{90}{Block-level Out-of-Order Scheduling}
           & \rotatebox{90}{Register File Hierarchy} \\ \specialrule{.1em}{.05em}{.05em}
    \textbf{CG-OoO}                         & \Checkmark  & \Checkmark  & \Checkmark  & \Checkmark  & \Checkmark  & \Checkmark & \Checkmark  & \Checkmark \\ \hline
    Braid~\cite{braid}                      & \Checkmark  &             & \Checkmark  &             &             & \Checkmark &             & \Checkmark \\ \hline
    WiDGET~\cite{widget}                    & \Checkmark  & \Checkmark  & \Checkmark  & \Checkmark  & \Checkmark  &            &             &            \\ \hline
    TRIPS~\cite{trips, burger200519}        & \Checkmark  &             & \Checkmark  &             & \Checkmark  & \Checkmark & \Checkmark  &            \\ \hline
    Multiscalar~\cite{multiscalar}          & \Checkmark  &             &             & \Checkmark  &             & \Checkmark & \Checkmark  &            \\ \hline
    CE~\cite{complexity}                    & \Checkmark  &             & \Checkmark  & \Checkmark  & \Checkmark  &            &             &            \\ \hline
    TP~\cite{trace}                         & \Checkmark  &             & \Checkmark  &             &             &            & \Checkmark  & \Checkmark \\ \hline
    MorphCore~\cite{morphcore}              &             & \Checkmark  &             & \Checkmark  &             &            & \Checkmark  &            \\ \hline
    BOLT~\cite{bolt}                        &             & \Checkmark  &             & \Checkmark  &             &            & \Checkmark  &            \\ \hline
    iCFP~\cite{icfp}                        &             &             & \Checkmark  & \Checkmark  &             &            &             &            \\ \hline
    ILDP~\cite{ildp}                        & \Checkmark  &             & \Checkmark  &             &             & \Checkmark &             & \Checkmark \\ \hline
    WaveScalar~\cite{wavescalar}            & \Checkmark  &             & \Checkmark  & \Checkmark  & \Checkmark  &            & \Checkmark  &            \\ \hline
    \end{tabular}
    \caption[Related Work: High Level Design Features Comparison]{Eight high
level design features of the CG-OoO architecture compared to the previous
literature.}
    \label{tab:rel_work1}
    \vspace {-5 pt}
    \end{center}
\end{table}

Multiscalar~\cite{multiscalar} evaluates a multi-processing unit capable of
steering coarse grain code segments, often larger than a basic-block, to its
processing units. It replicates register context for each computation unit,
increasing the data communication across its register files. TRIPS and
EDGE~\cite{edge,trips} are high-performance, grid-processing architectures that
uses static instruction scheduling in space and dynamic scheduling in time. It
uses Hyperblocks~\cite{hyperblock} to map instructions to the grid of
processors. Hyperblocks use branch predication to group basic-blocks that are
connected together through weakly biased branches. To construct Hyperblocks, the
TRIPS compiler uses program profiling.  While effective for improving
instruction parallelism, Hyperblocks lead to energy inefficient mis-speculation
recovery events. Palachara, et al.~\cite{complexity} supports a distributed
instruction window model that simplifies the wake-up logic, issue window, and
the forwarding logic. In this paper, instruction scheduling and steering is done
at instruction granularity. Trace Processors~\cite{trace} is an instruction flow
design based on dynamic code trace processing.  The register file hierarchy in
this work consists of several local register files and a global register file.
ILDP~\cite{ildp} is a distributed processing architecture that consists of a
hierarchical register file built for communicating short-lived registers locally
and long-lived registers globally. ILDP uses profiling and in-order scheduling
from each processing unit. In contrast to all of these proposals, the CG-OoO
compiler does not use program profiling (col. 5), and avoids static control
prediction by clustering instructions at basic-block granularity.  CG-OoO uses
local and segmented global registers to reduce data movement and SRAM storage
energy.

iCFP~\cite{icfp} addresses the head-of-queue\footnote{Stall of a ready operation
behind another stalling operation in a first-in-first-out (FIFO) queue} blocking
problem in the InO processor by building an execution model that, on every cache
miss, checkpoints the program context, steers miss-dependent instructions to a
side buffer enabling miss-independent instructions to make forward progress.
CFP~\cite{cfp} addresses the same problem in an OoO processor. Similarly,
BOLT~\cite{bolt}, Flea Flicker~\cite{fleaflicker}, and Runahead
Execution~\cite{mutlu2003runahead} are high ILP, high MLP\footnote{MLP: Memory
level parallelism.}, latency-tolerant architecture designs for energy efficient
out-of-order execution. All these architectures follow the runahead execution
model. BOLT uses a slice buffer design that utilizes minimal hardware resources.
CG-OoO solves the head-of-queue scheduling problem through a hierarchy of energy
efficient solutions including the Skipahead (Section~\ref{sec:issue_sch})
scheduler (col. 8).

WaveScalar~\cite{wavescalar} and SEED~\cite{nowatzki2015exploring} are
out-of-order data-flow architectures. The former focuses on solving the problem
of long wire delays by bringing computation close to data. The latter is a
complexity effective design that groups data-dependent instructions dynamically
and manages control-flow using \textit{switch instructions}.
MorphCore~\cite{morphcore} is an InO, OoO hybrid architecture designed to enable
single-threaded energy efficiency. It utilizes either core depending on the
program state and resource requirements. It uses dynamic instruction scheduling
to execute and commit instructions. In contract to the above, CG-OoO is a
single-threaded, block-level, energy efficient design that addresses the long
wire delays problem through clustering execution units, register files and
instruction queues close to one another. CG-OoO is end-to-end coarse-grain, and
code blocks do not need additional instructions to mange control flow.

\section{CG-OoO Architecture}\label{sec:exe_model}
The goal of the CG-OoO processor is to reach near the energy of the InO while
maintaining the performance level of OoO. This section introduces the CG-OoO as
a block-level execution model that leverages a hierarchy of solutions (software
and hardware) to save energy. Section~\ref{sec:cg_exe_eg} provides an execution
flow example. 

CG-OoO consists of multiple instruction queues, called Block Windows (BW), each
holding a dynamic basic-block and issuing instructions concurrently. BW's share
execution units (EU) to issue instructions (Figure~\ref{fig:eu_clusters}).
Several BW's and EU's are grouped to form execution clusters. CG-OoO uses
compiler support to group and statically schedule instructions. 

\subsection {Hierarchical Design}
\subsubsection{Hierarchical Architecture}
\begin{figure}
	\centering
	\includegraphics[width=\columnwidth]{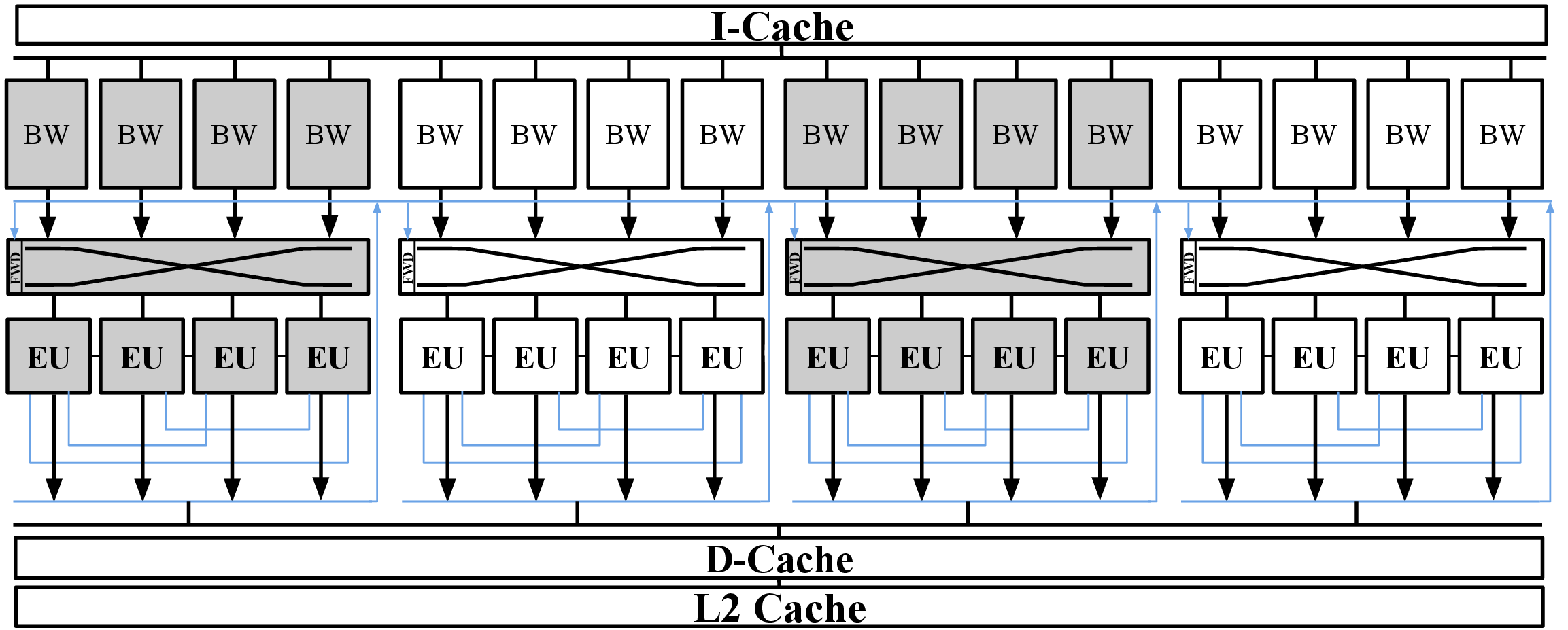} 
    \caption{The CG-OoO \{BW's, EU's\} cluster network.}
	\label{fig:eu_clusters}
\end{figure}

CG-OoO groups instructions into code-blocks that are fetched, dispatched, and
committed together. At runtime, each dynamic block is processed from a dedicated
BW. To manage data communication energy, BW and EU's are grouped together to
form clusters. Figure~\ref{fig:eu_clusters} shows CG-OoO clusters highlighted;
thin wires, in blue, enable data forwarding between EU's. Microarchitecture
clustering provides proportional energy-performance scaling based on program
load demands. Scalable architectures are previously studied by
\cite{fundamental,widget,chandrakasan1992low}. CG-OoO extends this concept to
energy efficient, block-level execution.

\subsubsection {Hierarchical Instruction Scheduling}
We use static instruction list scheduling on each basic-block to improve
performance and energy
\begin {enumerate*}[label=(\alph*)]
    \item by optimizing the schedule of predictable instructions along the
critical path,
    \item by improving MLP via hoisting memory operations to the top of
basic-blocks, and  
    \item by minimizing wasted computation due to memory mis-speculation
(Section~\ref{sec:mem_spec}). The compiler assumes L1-cache latency for memory
operations.
\end {enumerate*}

BW's in each cluster schedule instructions concurrently to hide each other's
head-of-queue stalls. We call this scheduling model block level parallelism
(BLP).  Furthermore, each BW supports a complexity effective, limited
out-of-order instruction issue model (Section~\ref{sec:issue_sch}) to address
unpredictable cases where coarse-grain scheduling cannot provide enough ILP.
These techniques combined help save energy by limiting the processor scheduling
granularity to the program runtime needs (Section~\ref{sec:cg_exe_eg} shows an
example).

\subsubsection {Hierarchical Register Files}
The CG-OoO register file hierarchy consists of: \textit{Global Register File
(GRF)}, and \textit{Local Register File (LRF)}. The GRF provides a small set of
architecturally visible registers that are dynamically managed while LRF is
statically managed, small, and energy efficient. The GRF is used for data
communication across BW's while LRF is used for data communication within each
BW. Each BW has its dedicated LRF. As shown in
Section~\ref{sec:reg_file_hier_energy}, 30\% of data communication
(register$\rightarrow$register and register$\leftrightarrow$memory) is done
through LRF's. To further save energy, the GRF is segmented and distributed
among BW's. GRF segmentation does not rely on a block-level execution model and
may be used independently. Similar register file models are studied
in~\cite{ildp,braid,trace,bmrf}. CG-OoO evaluates them from the energy
standpoint.

\subsection{Block-level Speculation}
\begin{figure}
	\centering
	\includegraphics[width=1.0\columnwidth]{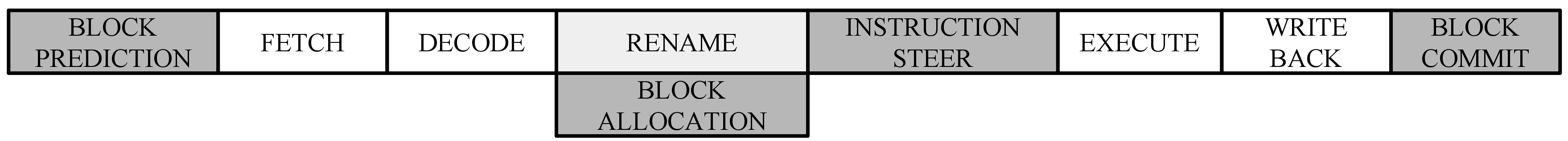} 
	\caption{CG-OoO processor pipeline stages}
	\label{fig:cg_ooo_pipeline}
\end{figure}

OoO processors avoid fetch stall cycles by performing BPU lookups immediately
before {\it{every fetch}} irrespective of the fetched instruction
types~\cite{EV8}; this leads to excessive speculation energy cost and redundant
BPU lookup traffic by non-control instructions which in turn may cause lower
prediction accuracy due to aliasing~\cite{mohammadidemand}. CG-OoO supports
energy efficient, block-level speculation by using only one BPU lookup per code
block. The compiler generates an instruction named \texttt{head} to
\begin{enumerate*}[label=(\alph*)]
    \item specify the start of a new code block,
    \item access the BPU to predict the next code block,
    \item trigger the Block Allocation unit to allocated a new BW and steer
upcoming instructions to it (Figure~\ref{fig:cg_ooo_pipeline}).
\end{enumerate*}
\texttt{head} is often ahead of its branch by at least one cycle making the
probability of front-end stall due to delayed branch prediction low.

\begin{figure}
	\centering
	\includegraphics[width=\columnwidth]{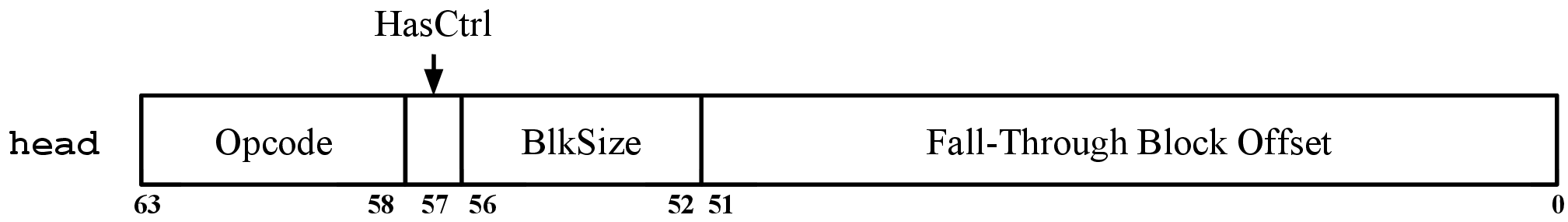} 
    \caption{The \texttt{head} instruction format}
	\label{fig:head_ins}
\end{figure}

\begin{figure}
	\centering
    \vspace {-10 pt}
	\includegraphics[width=\columnwidth]{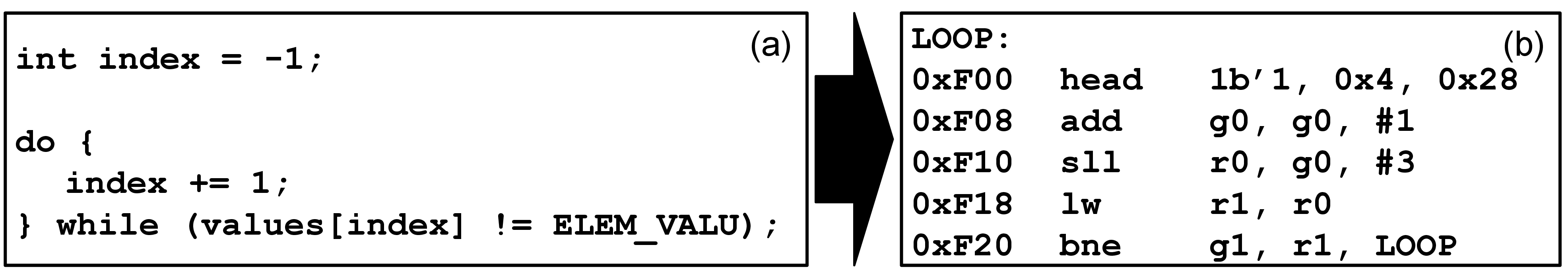} 
    \caption{A simple \texttt{do-while} loop program and its assembly code}
	\label{fig:simple_blk}
\end{figure}

Figure~\ref{fig:head_ins} shows the \texttt{head} instruction fields:
\begin {enumerate*}[label=(\alph*)]
    \item opcode,
    \item control instruction presence bit,
    \item block size\footnote{The compiler partitions code blocks larger than 32
instructions. Bird et al.~\cite{bird2007performance} shows the average size of
basic-blocks in the SPEC CPU 2006 integer and floating-point benchmarks are 5
and 17 operations respectively.},
    \item control instruction least significant address bits.
\end {enumerate*}
The example code in Figure~\ref{fig:simple_blk} shows \texttt{head} has
\texttt{HasCtrl=1'b1} indicating a control operation ends the basic-block. If
\texttt{HasCtrl=1'b0}, BPU lookup is disabled to save energy. In
Figure~\ref{fig:simple_blk}, local and global operands are identified by
\texttt{r} and \texttt{g} prefixes respectively.

\subsubsection{Squash Model}\label{sec:mem_spec}
CG-OoO supports block-level speculative control and memory squash. Upon control
mis-prediction, the front-end stalls fetching new instructions, all code blocks
younger than the mis-speculated control operation are flushed, and the remaining
code blocks are retired. The data produced by wrong-path blocks are
automatically discarded as such blocks never retire.  Once the BROB is empty,
the processor state is non-speculative, and it can resume normal execution. 

\subsection{CG-OoO Program Execution Flow\label{sec:cg_exe_eg}}
This section illustrates CG-OoO architecture with a code example. To better
understand the execution flow, Figure~\ref{fig:cg_ooo_pipeline} shows the CG-OoO
processor pipeline. The highlighted stages differ the traditional OoO. Control
speculation, dispatch, commit are at block granularity, and rename is only used
for global operands.  Section~\ref{sec:arch} discusses how each stage saves
energy.

\begin{figure*}
    \small
    \centering
    \begin{tabular}{@{}cc@{}}
        \includegraphics[width=.20\textwidth]{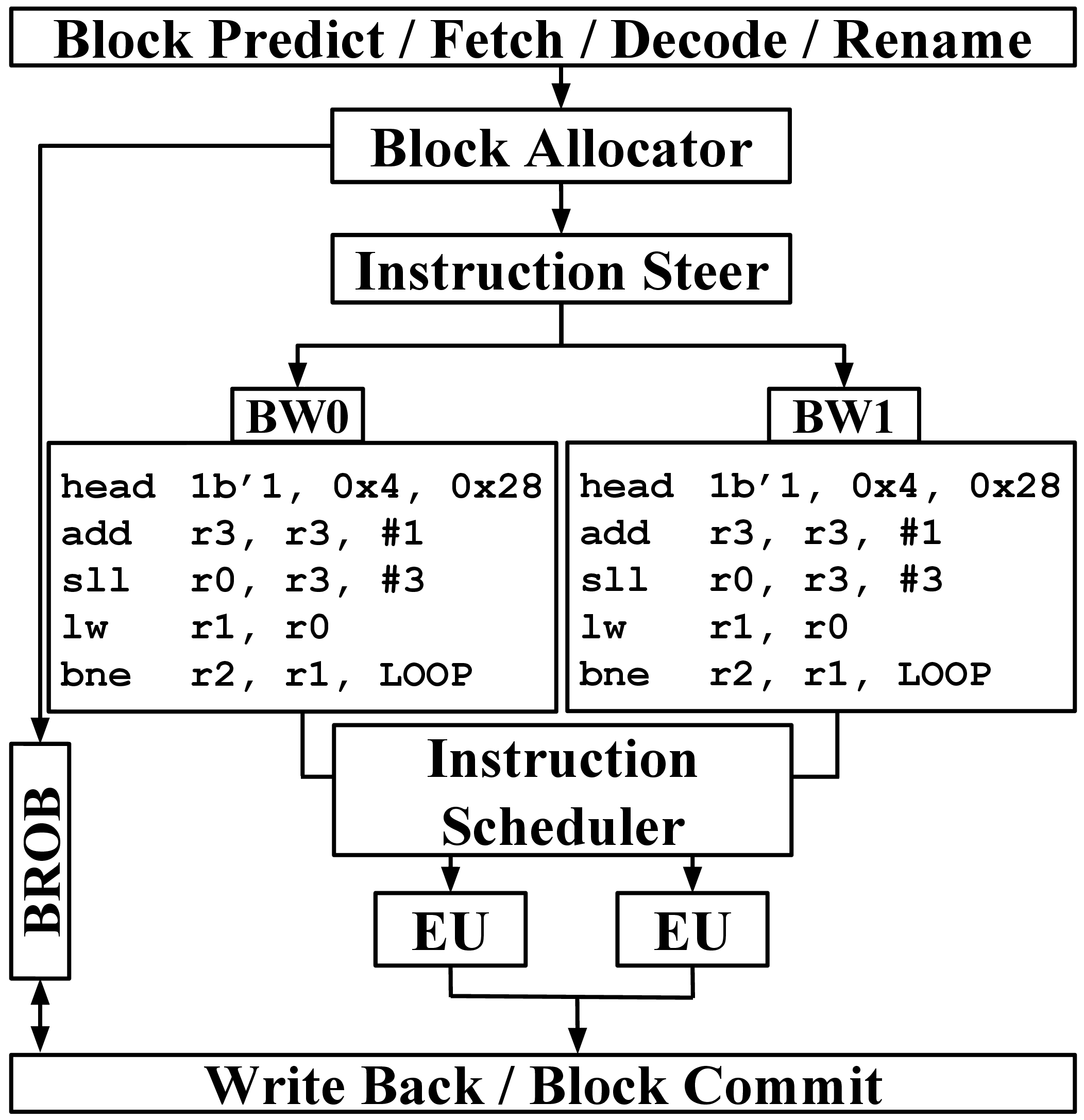} &
        \includegraphics[width=.80\textwidth]{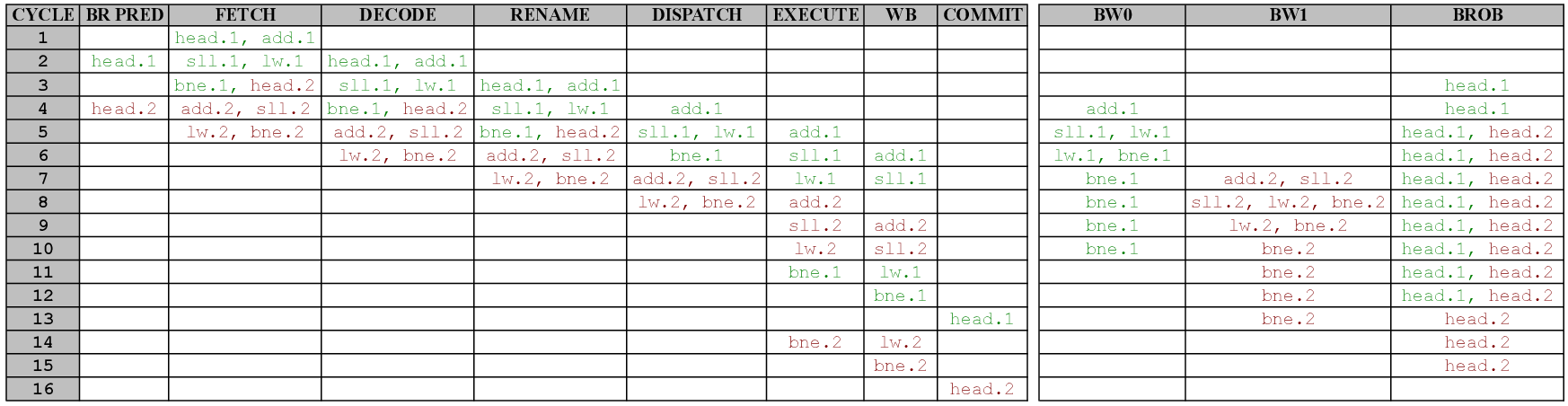} \\
    \end{tabular}
    \caption{(a) The CG-OoO architecture models. (b) Cycle-by-cycle flow of
example instructions in CG-OoO.}
    \label{fig:fetch_eg}
    \vspace {-10 pt}
\end{figure*}

Figure~\ref{fig:fetch_eg}a illustrates a two-wide superscalar CG-OoO. The
instruction scheduler issues one instruction per BW per cycle to the two EU's.
The code in BW's are two consecutive iterations of the abovementioned
\texttt{do-while} loop.  Figure~\ref{fig:fetch_eg}b shows the cycle-by-cycle
flow of instructions through the CG-OoO pipeline. Instructions in iterations 1
and 2 are green and red respectively. It also shows the contents of BW0, BW1,
and the Block Re-Order Buffer (BROB). Here, \texttt{lw} is a 4-cycle operation,
and all others 1-cycle.

In cycle 1, \texttt{\{head.1, add.1\}} instructions are fetched from the
instruction cache. In cycle 2, the immediate field of \texttt{head.1} is
forwarded to the BPU. In cycle 3, \texttt{head.1} speculates the next code block
\textit{before} the control operation, \texttt{bne.1}, is fetched; furthermore,
the Block Allocator assigns BW0 to the instructions following \texttt{head.1},
and BROB reserves an entry for \texttt{head.1} to stores the runtime status of
its instructions. In cycle 4, BW0 receives its first instruction. In cycle 5,
\texttt{add.1} is issued while more instructions join BW0. In cycle 10, the last
instruction of iteration 1 leaves BW0. In cycles 11, BW0 is available to hold
new code blocks. In cycle 13, \texttt{head.1} is retired as all its instructions
complete execution; at this point, all data generated by the block operations
will be marked non-speculative.

%

%
%

\section{CG-OOO Micro-architecture}\label{sec:arch}
This sections presents the CG-OoO pipeline micro-architecture details and
highlights their energy saving attributes. These stages save energy by utilizing
several complexity effective techniques through
\begin{enumerate*}[label=(\alph*)]
    \item the use of small tables,
    \item reduced number of table accesses, and
    \item hardware-software hybrid instruction scheduling.
\end{enumerate*}

\subsection{Branch Prediction}
\begin{figure}[h]
	\centering
	\includegraphics[width=1.0\columnwidth]{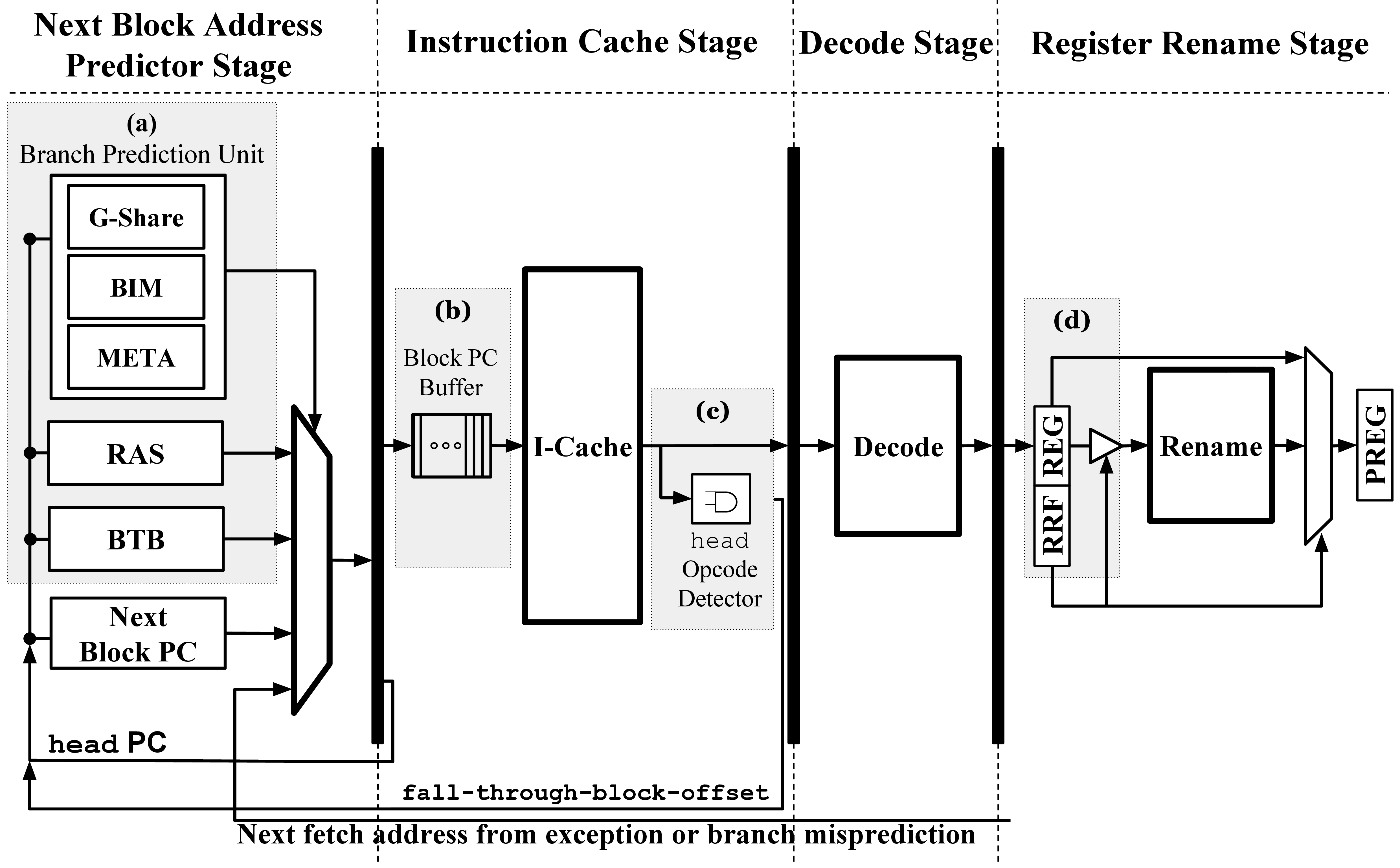} 
    \caption{The Branch Prediction Unit (BPU) micro-architecture.}
	\label{fig:bpu_model}
\end{figure}

Figure~\ref{fig:bpu_model}a shows the micro-architectural details of the branch
prediction stage in the CG-OoO processor; it consists of the Branch Predictor
(BP)~\cite{EV8}, Branch Target Buffer (BTB), Return Address Stack (RAS), and
Next Block-PC. Equation~\ref{eqn:next_blk_pc} shows the Next Block-PC
computation relationship.

\begin{equation}
    PC_{Next-\texttt{head}} = PC_{\texttt{head}} + \texttt{fall-through-block-offset}
    \label{eqn:next_blk_pc}
\end{equation}

The \texttt{fall-through-block-offset} is the immediate field of the
\texttt{head} instruction shown in Figure~\ref{fig:head_ins}. In the CG-OoO
model, only \texttt{head} PC's access the BPU.  Upon lookup, a \texttt{head} PC
is used to predict the next \texttt{head} PC. Speculated PC's are pushed into a
FIFO queue, named \textit{Block PC Buffer}, dedicated to communicate block
addresses to Fetch (Figure~\ref{fig:bpu_model}b).

Once completed, each control operation verifies the next-block prediction
correctness and updates the corresponding BPU entry(ies) accordingly. Since the
BPU is indexed by \texttt{head} PC's, control operations access their
corresponding BPU entry(ies) by computing their \texttt{head} PC using
Equation~\ref{eqn:head_pc}.

\begin{equation}
    PC_{\texttt{head}} = PC_{control-op} - \texttt{code-block-offset}
    \label{eqn:head_pc}
\end{equation}

\subsection{Fetch Stage}
\begin{figure}[h]
	\centering
	\includegraphics[width=\columnwidth]{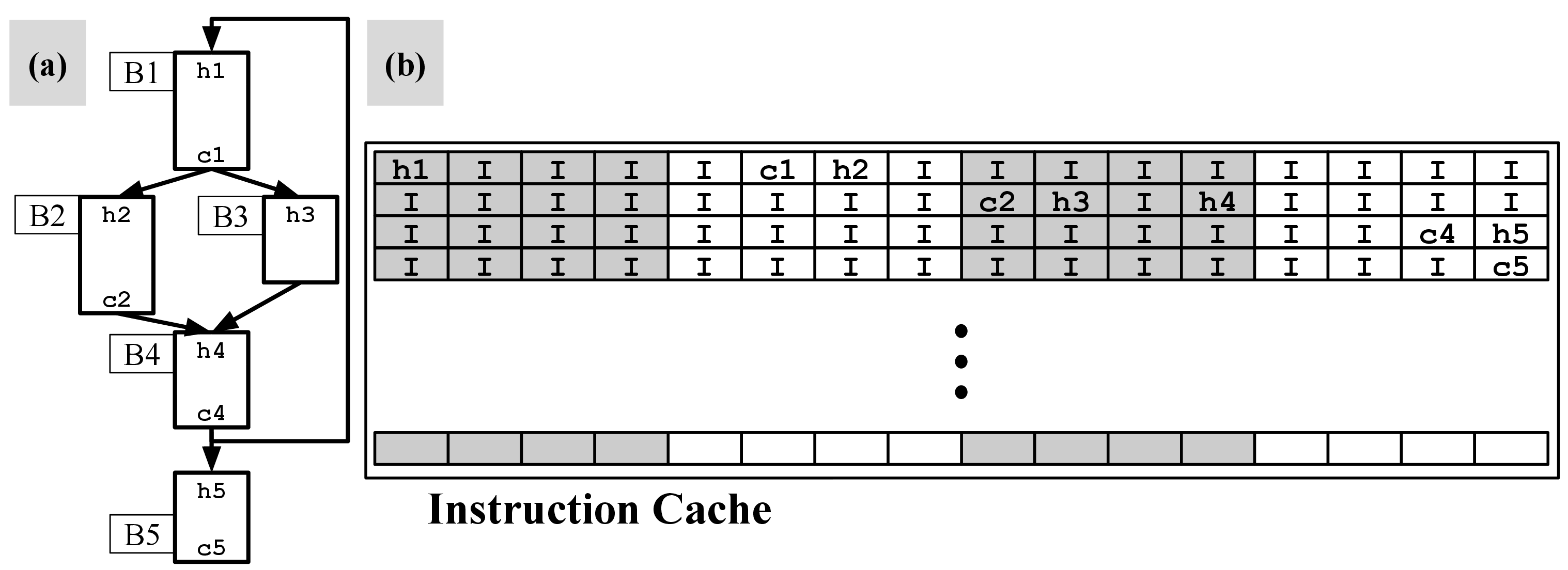} 
    \caption{(a) A control flow graph (CFG) with 5 blocks labeled B1 to B5. (b)
Mapping of CFG operations onto a 4-wide instruction cache.}
	\label{fig:blk_ins_cache}
\end{figure}

Figure~\ref{fig:blk_ins_cache}a illustrates a control flow graph with five
basic-blocks. Each block is marked with its \texttt{head} identifier,
\texttt{h}, at the top, and its control operation identifier (if any),
\texttt{c}, at the bottom.  Figure~\ref{fig:blk_ins_cache}b illustrates the
mapping of these basic-blocks to the instruction cache where each box represents
an instruction and each set of adjacent boxes with the same color represent a
fetch group.\footnote{This section assumes no
fetch-alignment~\cite{mahin1997superscalar}.} Entries marked \texttt{I}
represent non-control, non-\text{head} operations in each basic-block in
Figure~\ref{fig:blk_ins_cache}a. As shown in Figure~\ref{fig:blk_ins_cache}b, an
arbitrary number of \texttt{head}'s may exist in a fetch group. Fetch and BPU
handle all cases.

The Block PC Buffer holds either a next-PC address or a \texttt{64'b0} where the
former is the next code block to fetch from the cache, and the latter is a hint
that next-PC is unknown. An unknown PC happens when a \texttt{head} operation,
\texttt{hh}, is predicted \textit{not-taken} and \texttt{hh} itself is not yet
fetched. Recall, the predictor needs to have the
\texttt{fall-through-block-offset} of \texttt{hh} to predict the next block.  In
such cases, Fetch assumes the fall-through block is adjacent to the \texttt{hh}
block in memory; so, it continues fetching the next block while the fall-through
block address for \texttt{hh} is computed.

\subsection{Decode \& Register Rename Stages}
The Decode micro-architecture follows that of the conventional OoO except for
its additional functionality to identify global and local register operands by
appending a 1-bit flag, named \textit{Register Rename Flag (RRF)}, next to each
register identifier. If an instruction holds a global operand, it accesses the
register rename (RR) tables for its physical register identifier; otherwise, it
would skip RR lookup (Figure~\ref{fig:bpu_model}d). Skipping the register rename
stage reduces the renaming lookup energy by 30\% on average. This saving is
realized due to our block-level execution model. Our RR evaluations use the
Merged Rename and Architectural Register File model discussed
in~\cite{kessler1998alpha, mips, hinton2001microarchitecture}. 

\subsection{Issue Stage}
Before discussing the CG-OoO issue model, let us visit the Block Window
micro-architectural shown in Figure~\ref{fig:block_window}. It consists of an
Instruction Queue (IQ), a Head Buffer (HB), a dedicated LRF, a GRF segment, and
a number of EU's. IQ is a FIFO that holds code block instructions. HB is a
\textit{small buffer} that holds instructions waiting to be issued by the
Instruction Scheduler in a content accessible memory (CAM) array. HB pulls
instructions from the IQ and waits for their operands to become ready for issue.
The CG-OoO issue model allows register file accesses only to operations in the
HB thereby
\begin{enumerate*}[label=(\alph*)]
    \item avoiding the OoO post-issue register file read cycle, and 
    \item saving the pre-issue large data storage
overhead~\cite{gonzalez2010processor} by only storing operands in HB's.
\end{enumerate*}
Because the number of operations in all HB's is a fraction of all in-flight
instructions, this model is as fast as the OoO pre-issue model, and more energy
efficient than both models.


\subsubsection{Skipahead Instruction Scheduler\label{sec:issue_sch}} 
\begin{figure}
\centering
    \begin{subfigure}[b]{0.35\columnwidth}
        \includegraphics[width=\columnwidth]{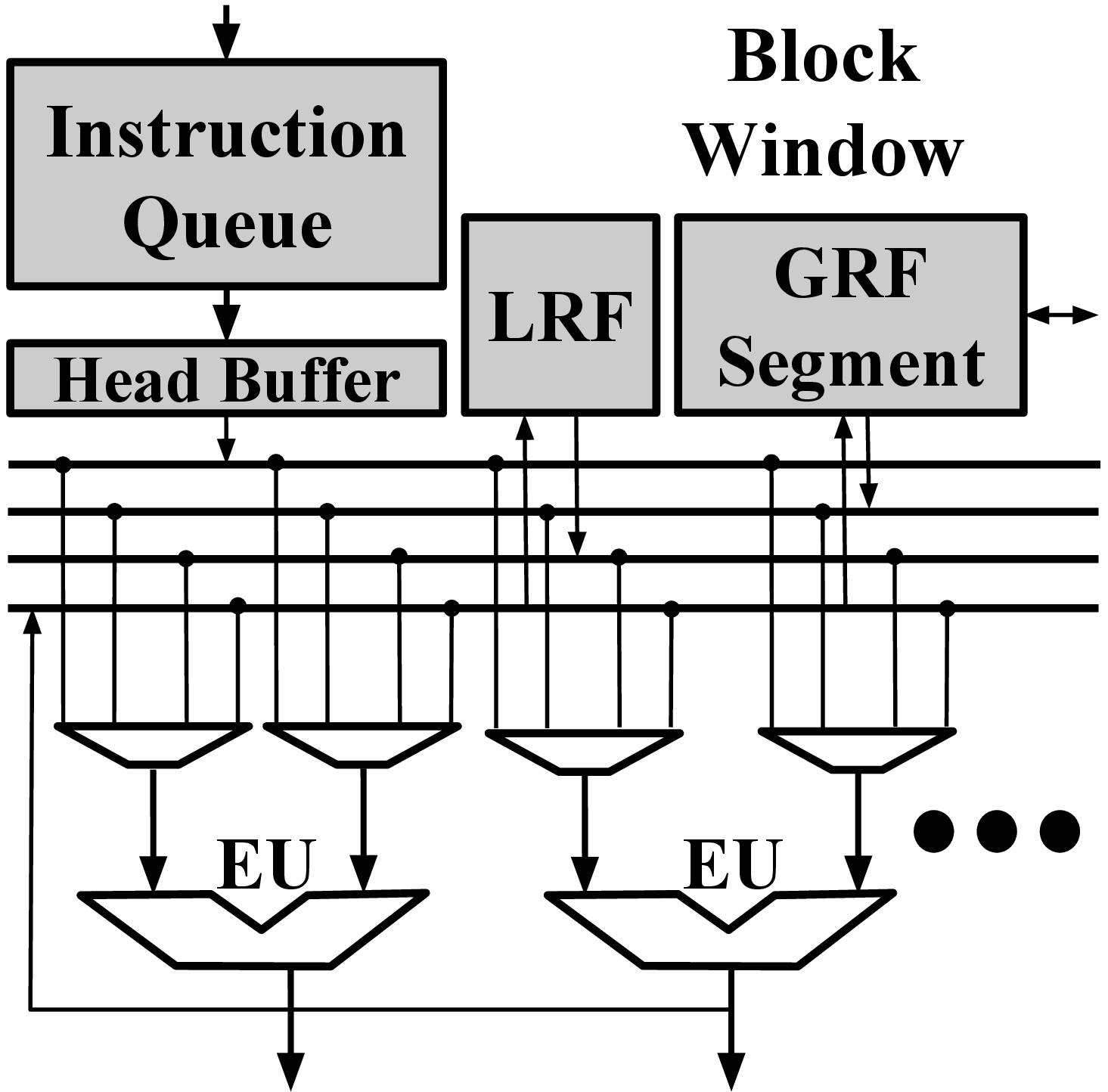} 
        \caption{The Block Window (BW) microarchitecture}
        \label{fig:block_window}
    \end{subfigure}
    ~
    \begin{subfigure}[b]{0.61\columnwidth}
        \includegraphics[width=\columnwidth]{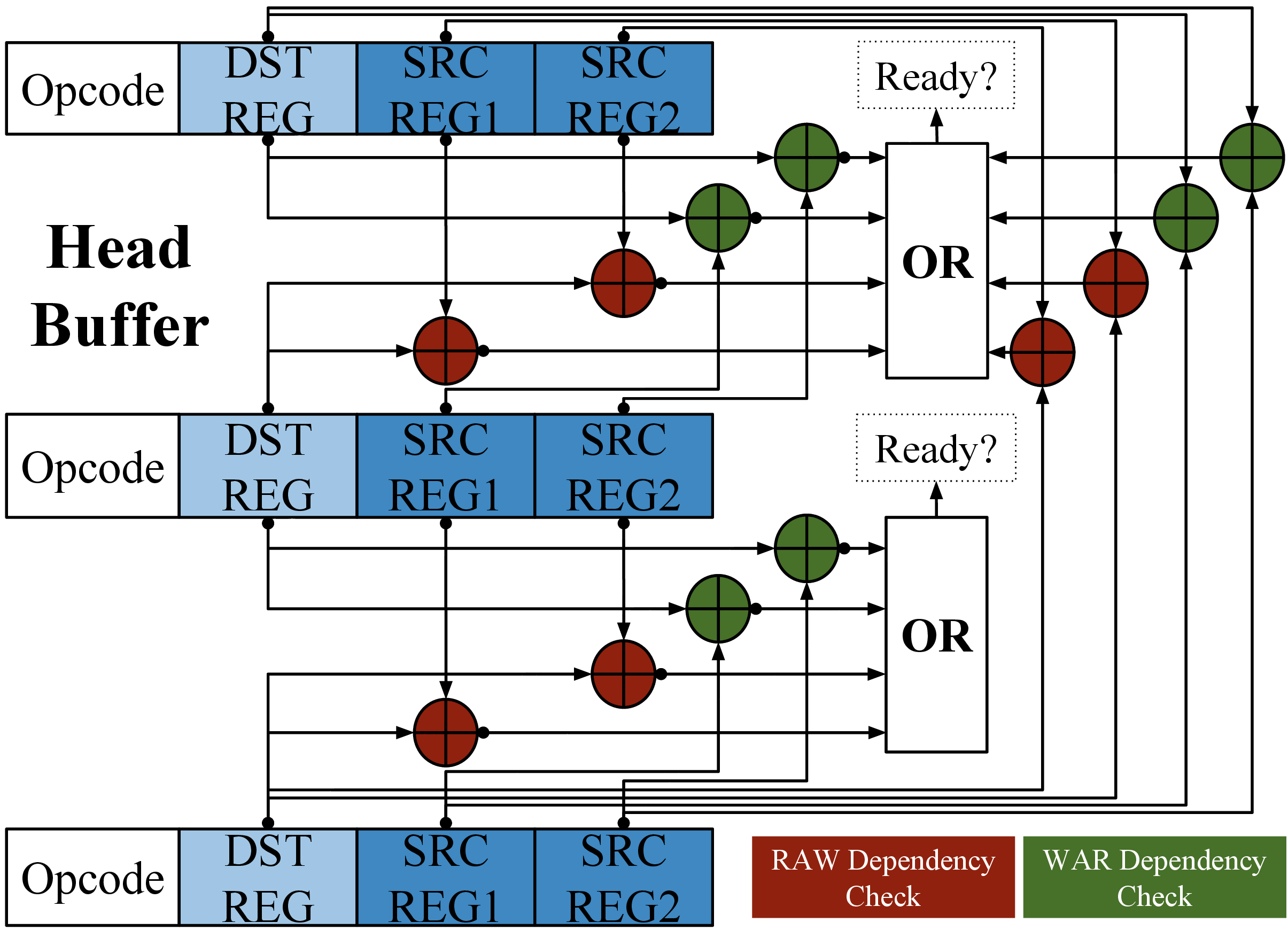} 
         \caption{The logic to support the Skipahead issue model in Head Buffer (HB)}
        \label{fig:head_buffer}
    \end{subfigure}
    \caption{}
\end{figure}

The Skipahead model allows \textit{limited out-of-order} issue of operations.
The term \textit{limited} means out-of-order instruction issue is restricted to
only the HB instructions, a subset of all code block instructions. When a HB
instruction, \textit{Ins}, becomes ready prior to HB instruction(s) ahead of it,
if \textit{Ins} does not create a true or false dependency with older
instructions, it may be issued out-of-order. Figure~\ref{fig:head_buffer} shows
the complexity effective XOR logic used for dependency checking.

For example, assuming a three-entry HB, in Figure~\ref{fig:issue_model_eg},
instructions are issued as \texttt{\{1, 3\}} followed by \texttt{\{2, 4\}}.
Before issuing \texttt{3}, its operands are dependency checked against those of
\texttt{2}.

The Skipahead model improves the CG-OoO performance by 41\%
(Section~\ref{sec:perf}) while enabling a selection and wakeup model no more
energy hungry than an in-order issue model. The wakeup unit presents three
sources of energy efficiency.
\begin{enumerate*}[label=(\alph*)]
    \item In each BW, the wakeup unit uses a small HB storage space to hold
operand \textit{data}. In contrast to OoO, operations stored in the Instruction
Queue are not included in the wakeup process.
    \item The wakeup unit searches small CAM tables for \textit{source
operands}. For instance, in a CG-OoO processor with 8 BW's, each with 3 HB
entries, the wakeup unit accesses 48 CAM source operand entries. The OoO
baseline assumes 128~\cite{mutlu2003runahead,coorporation2009intel} in-flight
operations in Instruction Window to search for ready operands.
    \item Local write operands wakeup source operands associated with their own
BW \textit{only}.
\end{enumerate*}

\subsection{Memory Stage}
\begin{figure}
	\centering
	\includegraphics[width=0.5\columnwidth]{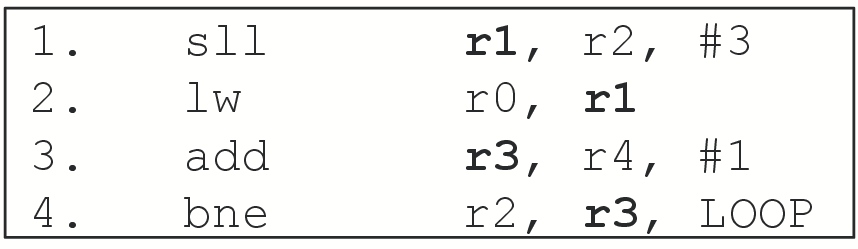} 
    \caption{A code snippet with two data-dependencies.}
	\label{fig:issue_model_eg}
\end{figure}

The CG-OoO Memory stage consists of a load-store-unit (LSU) that operates at
instruction granularity. A squash is triggered when a \texttt{sw} conflicts with
a younger \texttt{lw} at which point the block holding the \texttt{lw} is
flushed; this means {\it{useful}} instructions older than \texttt{lw} are also
squashed. For instance, in Figure~\ref{fig:issue_model_eg}, if operation 2 were
to trigger a memory mis-speculation event, the entire block, including operation
1, would be squashed. Flushing useful operations is called \textit{wasted
squash} which the compiler reduces by hoisting memory operations toward the top
of basic-blocks. Efficient memory speculation models such as
NoSQ~\cite{sha2006nosq} can further improve processor energy efficiency by
replacing associative LSU lookups with indexed lookups. Evaluating the energy
impact of such designs is outside of the scope of this work.

\subsection{Write Back \& Commit Stage}
Figure~\ref{fig:brob_entry} shows the contents of a BROB entry; it holds the
block sequence number, block size, and block global write (\texttt{GW}) register
operand identifiers. \texttt{BlkSize} is initialized by the corresponding
\texttt{head} operation. The \texttt{GW} fields are updated by instructions with
global write registers as they are steered from the Register Rename stage to
their BW. The compiler controls the number of global write operands per code
block.

\subsubsection{Write-Back Stage}
\begin{figure}
	\centering
    \vspace {-10 pt}
	\includegraphics[width=1.0\columnwidth]{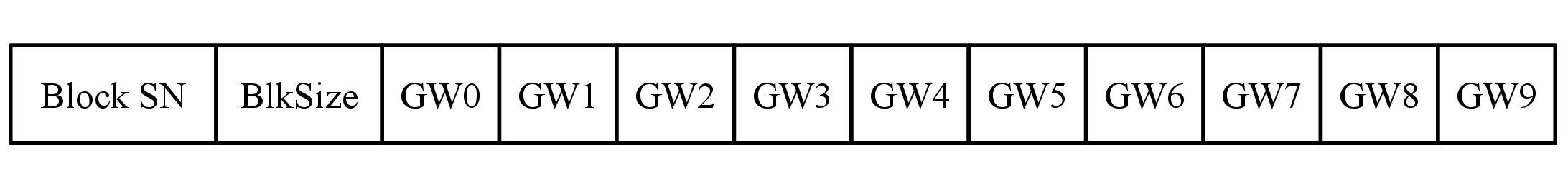} 
    \caption{A Block Re-Order Buffer (BROB) entry}
	\label{fig:brob_entry}
\end{figure}

Once an instruction completes, it writes its results into either a designated
register file entry (global or local) or into the store queue. In
Figure~\ref{fig:brob_entry}, \texttt{BlkSize} is decremented upon each
instruction complete; once its value is zero, the corresponding \textit{block}
is \textit{completed}.

\subsubsection{Commit Stage}
A block is committed when it is \textit{completed} and is at the head of the
BROB. During commit, all global registers modified by the block are marked
\textit{Architectural} using the \texttt{GW} fields in BROB. Upon commit,
\texttt{sw} operations in the committing code block retire; in doing so, the
Store-Queue ``commit" pointer moves to the youngest \texttt{sw} belonging to the
committing block. This \texttt{sw} is found via searching for the youngest store
operation whose \texttt{Block SN} matches that of the committing block. Note,
our LSU holds a \texttt{Block SN} column.

Checkpoint-based processors~\cite{checkpoint,cristal2002large,cristal2004out}
propose a general concept applicable to many architectures (e.g.
iCFP\cite{icfp}). While outside of the scope of this work, coarse-grain
checkpoint-based processing is promising for extending the energy efficiency of
CG-OoO.

\subsection{Squash Handling}
CG-OoO handles squash events through the following steps:
\begin{enumerate*}[label=(\alph*)]
    \item the BPU history queue and Block PC Buffer flush the content
corresponding to wrong-path blocks. The code block PC resets to the start of the
right path; in case of a control mis-speculation, the right path is the opposite
side of the control operation, and in case of a memory mis-speculation, it is
the \textit{start} of the same code block.

    \item All BW's holding code blocks younger than the mis-speculated operation
flush their IQ, Head Buffer, and mark LRF registers invalid.

    \item LSU flushes operations corresponding to the code block younger
than the mis-speculated operation by comparing the mis-speculated \texttt{Block
SN} against that of memory operations.

    \item BROB flushes code block entries younger than the mis-speculated
operation. The remaining blocks complete execution and commit.
\end{enumerate*}

\section{Methodology}\label{sec:method}
The evaluation setup consists of an in-house compiler, simulator and energy
model. The compiler performs \textit{Local Register Allocation} and
\textit{Global Register Allocation} as well as \textit{Static Block-Level List
Scheduling} for each program basic-block. This means the output ISA differs from
the \textit{gcc} generated x86 ISA. The simulator consists of a Pin-based
functional emulator attached to a timing simulator~\cite{pin}. The emulator
supports wrong-path execution. The dynamic instructions produced by the emulator
are mapped to an internal ISA for processing by the timing simulator.
Table~\ref{tab:core_params} outlines the configurations used by the simulator to
support the timing and energy model for the CG-OoO, OoO, and InO processors. The
simulator uses the cache and memory model in~\cite{das2015slip}. All evaluations
support instruction fetch alignment. They also support data-forwarding between
EU's. Evaluations start after 2 billion instructions, warm-up for 2 million
instructions, and simulate the following 20 million instructions.  The simulator
handles precise exceptions by executing instructions in its in-order mode. Once
recovered from the exception, the program resumes normal execution. 

\begin{table}[htb]
    \centering
    \small
    \vspace {-5 pt}
    \begin{tabular}{l|l}
        \hline
        \multicolumn{2}{c}{{\bf Shared Parameters}} \\ \hline
        ISA				                   & x86\\ \hline
        Technology Node				       & 22nm \\ \hline
        System clock frequency		       & 2GHz\\ \hline
        L1, assoc, latency                 & 32KB, 8, 4 cycles\\
        L2, assoc, latency                 & 256KB, 8, 12 cycles\\ 
        L3, assoc, latency                 & 4MB, 8, 40 cycles\\ \hline
        Main memory latency                & 100\\ \hline
        Instruction Fetch Width		       & 1 to 8 \\ \hline
        Branch Predictor	               & Hybrid\\
        \ - G-Share, Bimodal, Meta	       & 8Kb, 8Kb, 8Kb\\
        \ - Global Pattern Hist	           & 13b\\
        BTB, assoc, tag	                   & 4096 entries, 8, 16b\\ \hline
        Load/Store Queues                  & 64, 32 entries, CAM \\

        \hline
        \multicolumn{2}{c}{{\bf InO Processor}} \\ \hline
        Pipeline Depth                 & 7 cycles \\
        Instruction Queue              & 8 entries, FIFO \\
        Register File                  & 70 entries (64bit) \\
        Execution Unit                 & 1-8 wide \\
        
        \hline
        \multicolumn{2}{c}{{\bf OoO Processor}} \\ \hline
        Pipeline Depth                 & 13 cycles \\
        Instruction Queue              & 128 entries, RAM/CAM \\
        Register File                  & 256 entries (64bit) \\
        Execution Unit                 & 1-8 wide \\
        Re-Order Buffer                & 160 entries \\
        
        \hline
        \multicolumn{2}{c}{{\bf CG-OoO Processor}} \\ \hline
        Pipeline Depth                 & 13 cycles \\
        Number of BW's                 & 3-18 \\
        Instruction Queue / BW         & 10 entries, FIFO \\
        Head Buffer / BW               & 2-5 entries, RAM/CAM \\
        Execution Unit / BW            & 1-8 wide \\
        GRF, LRF / BW                  & 256, 20 entries (64bit) \\
        GRF Segments                   & 1-18 \\
        Number of Clusters             & 1-3 clusters \\
        Block Re-Order Buffer          & 16 entries \\
        \hline
    \end{tabular}
    \caption{System parameters for each individual core}
    \label{tab:core_params}
    \vspace {-10 pt}
\end{table}

\subsection{Energy Model}
Our energy model produces per-access energy numbers for the simulator to use to
compute the total energy of each hardware unit. This model extends the energy
model in~\cite{das2015slip} to support tables, caches, wires, stage registers,
and execution unit energies and areas. It estimates per-access dynamic energy
and per-cycle static energy consumption. The simulator computes the
\textit{total dynamic energy} by incrementing per-access energy of each unit. It
computes the \textit{total static energy} by multiplying the number of
simulation cycles by the per-cycle leakage energy of each unit. Other logical
blocks in the processor (e.g. control modules) are assumed to have similar
energy costs for the baseline OoO and the CG-OoO, and to have secondary effect
on the overall energy difference. 

RAM tables are modeled as standard SRAM units accessed through decoder and read
through sense amplifiers. Static and dynamic energy are generated using SPICE.
Then, additional steps including area estimation, energy scaling for different
port configurations and cache structures are done. Similarly, CAM tables are
designed as standard SRAM units accessed through a driver input module and read
through sense amplifiers. To evaluate the energy and area of pipeline stage
registers, 6-NAND gate positive edge-triggered flip-flops (FF) are simulated in
SPICE.

Different 64-bit execution units including the add, multiply, divide units for
arithmetic and floating-point operations are developed in Verilog and simulated
in the Design Compiler~\cite{azizi2010energy}. The Design Compiler provides
per-operation energy numbers for each unit.

HotSpot~\cite{huang2006hotspot} is used for optimal chip floorplan and wire
length optimization. To extract wire energy numbers, we upgraded HotSpot to
report wire energy using its wire length outputs. The energy per access used for
wires is 0.08 pJ/b-mm at the 22nm technology node~\cite{cao2002predictive}. The
simulator assumes all wires have 0.5 activity factor; so, every time the
simulator drives a wire, its energy consumption is incremented by half of its
per-access energy.

\section{Evaluation}\label{sec:eval}
CG-OoO achieves the performance of OoO at 48\% of its total energy cost on SPEC
Int 2006 benchmarks~\cite{spec}. This section quantifies the performance and
energy benefits of the CG-OoO processor and the pipeline stages that contribute
to its superior energy profile.

\subsection{CG-OoO Performance Analysis}\label{sec:perf}
\begin{figure}[h]
\centering
    \begin{subfigure}[b]{\columnwidth}
        \centering
        \includegraphics[width=1.0\columnwidth]{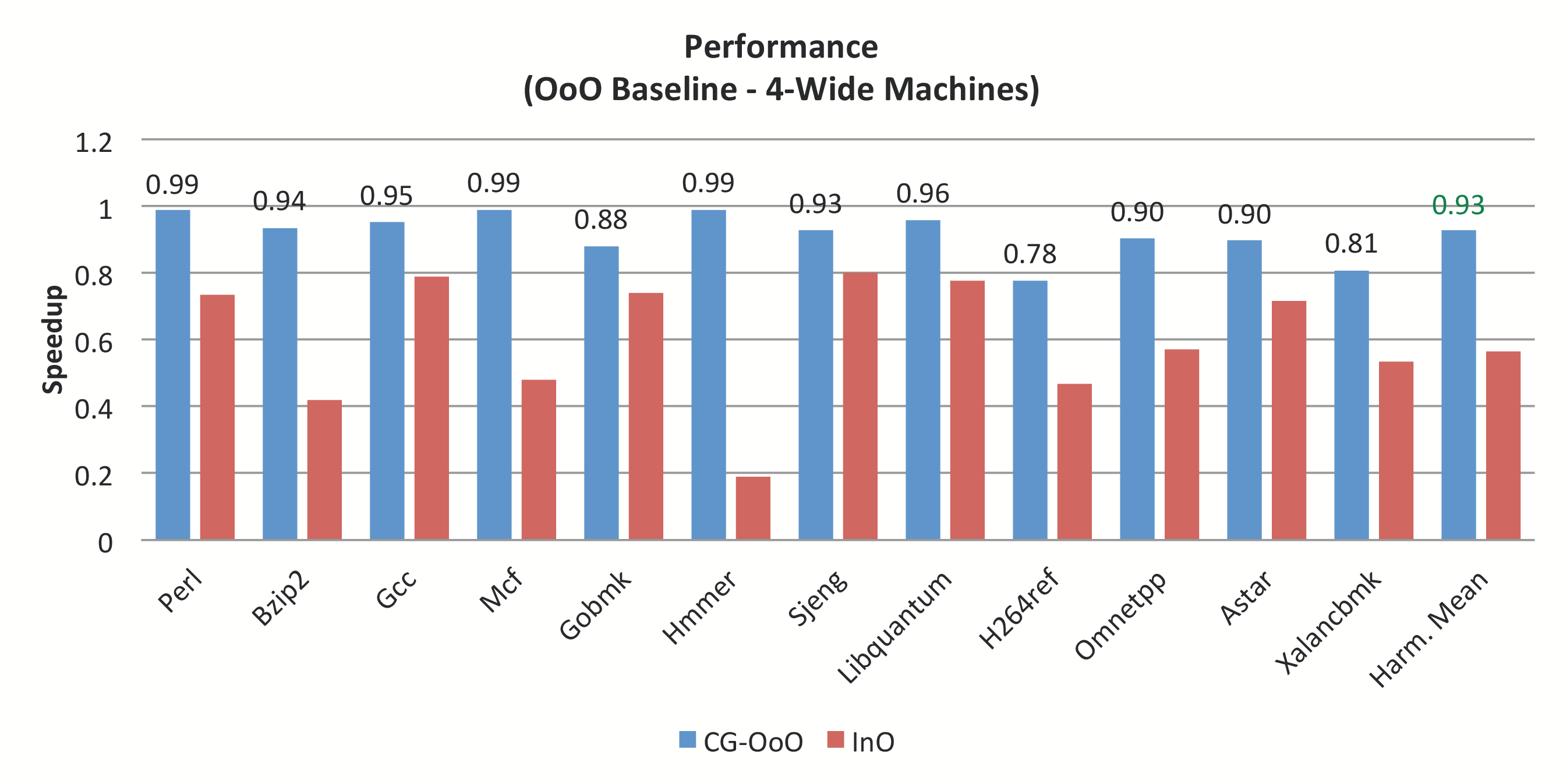}
        \caption{CG-OoO with 4-wide front-end \& back-end}
        \label{fig:perf_4w}
    \end{subfigure}

    \begin{subfigure}[b]{\columnwidth}
        \includegraphics[width=1.0\columnwidth]{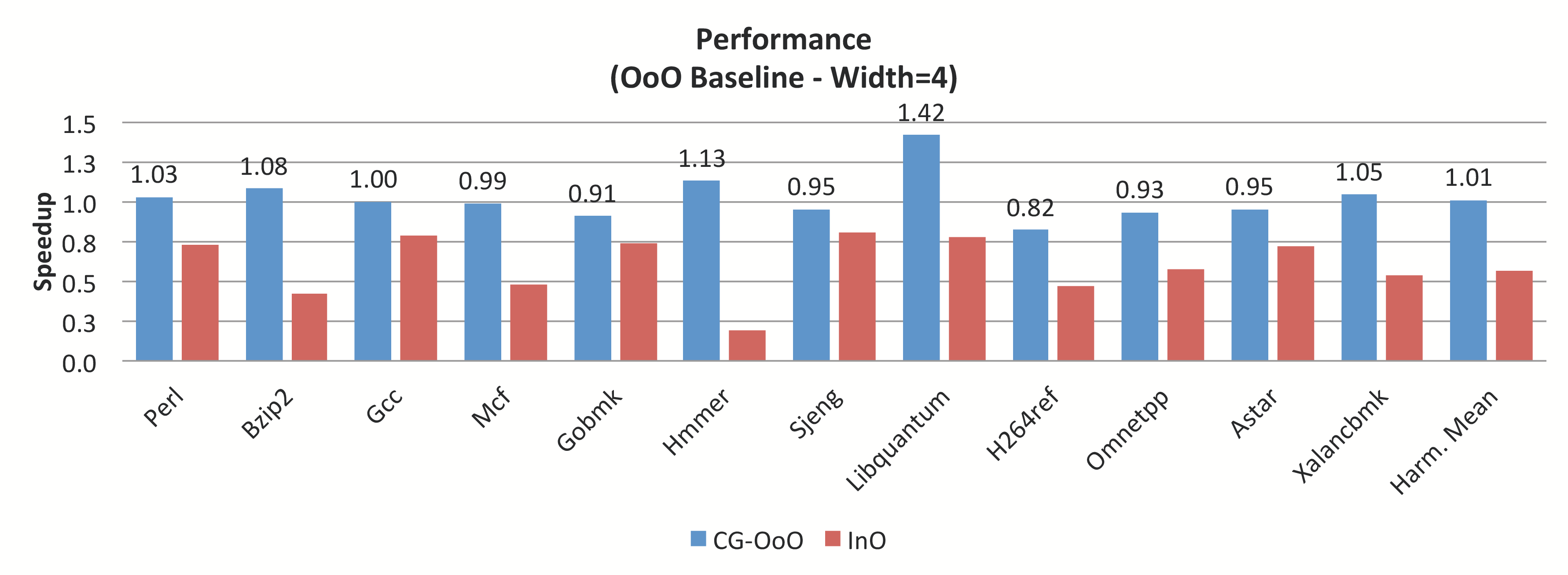}
        \caption{CG-OoO with 4-wide front-end}
        \label{fig:perf}
    \end{subfigure}

    \caption{Performance of InO and CG-OoO normalized to OoO.}
\end{figure}

Figure~\ref{fig:perf_4w} uses a 4-wide OoO superscalar processor as the baseline
for illustrating the relative performance of a 4-wide InO processor with a
CG-OoO processor (4-wide front-end and 4 EU's arranged as a single cluster). In
this case, the CG-OoO harmonic mean performance is 7\% lower than the OoO
baseline. Performance results are measured in terms of instructions per cycle
(IPC). In Figure~\ref{fig:perf}, the same 4-wide InO and OoO configurations are
compared against a CG-OoO model with a 4-wide front-end and 12 EU's
spread across 3 clusters. In this configuration, the CG-OoO ILP reaches that of
the OoO. As can be observed for Hmmer, Bzip2, and Libquantum benchmarks, the
higher availability of computation resources allows exploiting higher ILP.

\begin{figure}
	\centering
    \vspace {-10 pt}
	\includegraphics[width=1.0\columnwidth]{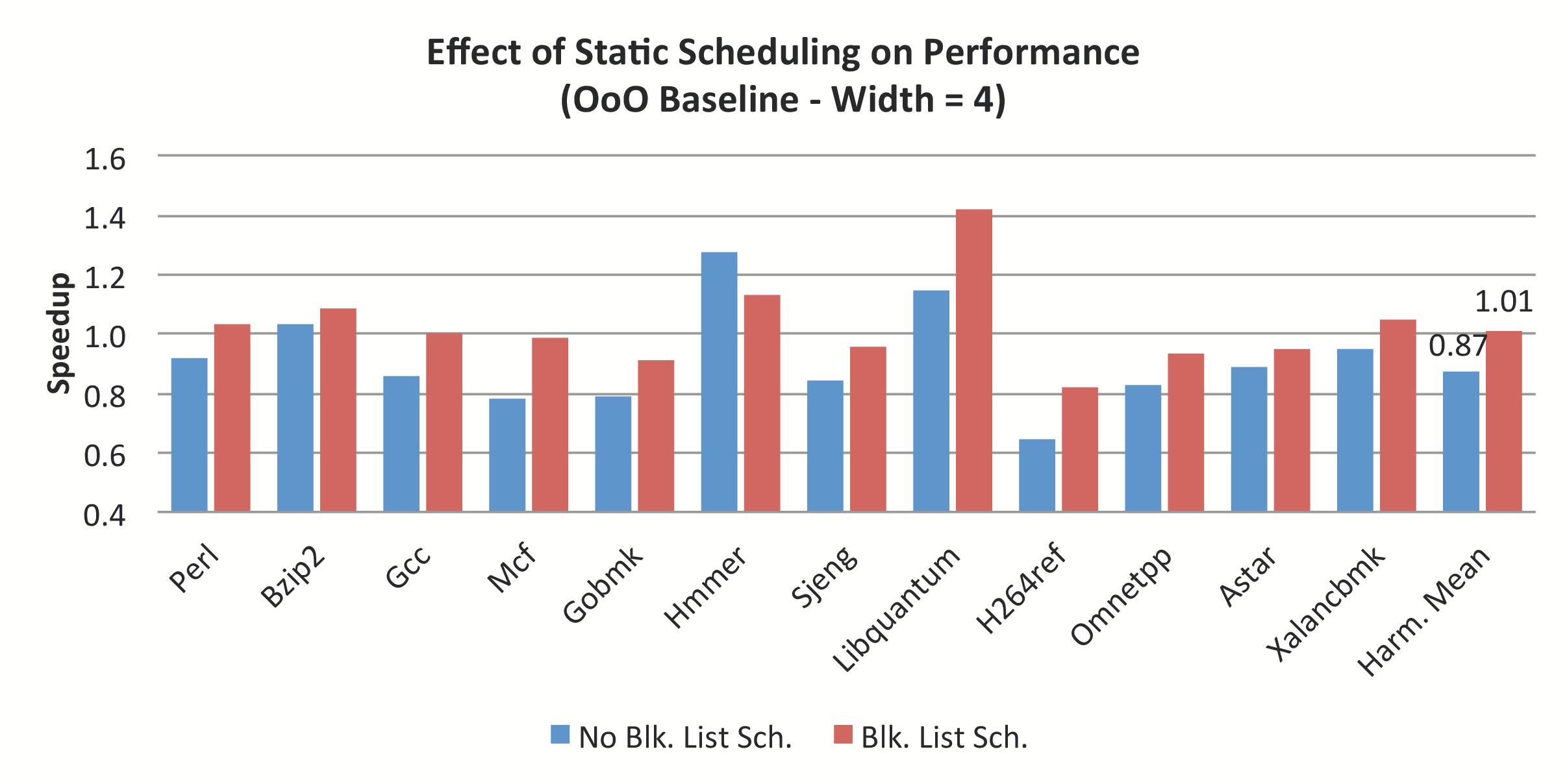}
    \caption{Effect of static block-level list scheduling on CG-OoO performance.
The Skipahead 4 dynamic scheduling model is used for both results (see
Figure~\ref{fig:skipahead}).}
	\label{fig:perf_static_sch}
\end{figure}

The first source of performance gain is \textit{static block-level list
scheduling}. Figure~\ref{fig:perf_static_sch} shows the effect of static
scheduling on performance. On average, static scheduling increases the CG-OoO
performance by 14\%. In case of Hmmer, 19\% more MLP is observed with the
original binary schedule generated using \textit{gcc} (with -O3 optimization
flag) than the code generated using static block-level list scheduling. The
higher MLP is due to a superior global code schedule which in turn leads to
fewer stall cycles. In both cases, Hmmer performs better than the OoO baseline.

\begin{figure}[h]
	\centering
	\includegraphics[width=1.0\columnwidth]{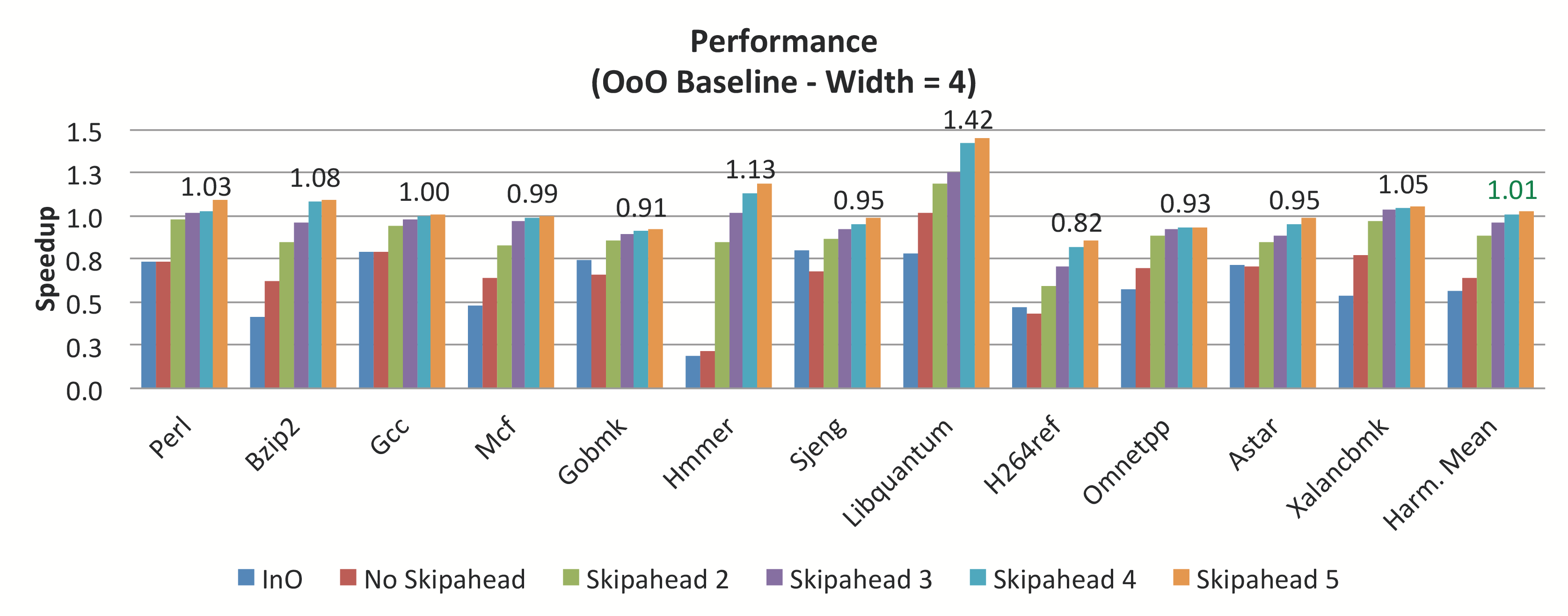}
    \caption{The performance attribute of the Skipahead model.}
	\label{fig:skipahead}
\end{figure}

The next source of performance gain is through BLP.  To illustrate the
contribution of BLP, let us assume each BW can issue up to four operations in-order;
that is, if an instruction at the head of a BW queue is not ready to issue,
younger, independent operations in the same queue do not issue. Other BW's,
however, can issue ready operations to hide the latency of the stalling BW. The
\textit{No Skipahead} bar in Figure~\ref{fig:skipahead} refers to this setup. It
shows on average 17\% of the performance gap between the InO and OoO is closed
through BLP.  Benchmarks like H264ref and Sjeng exhibit better performance for
the InO model. This is because the InO processor has a shallower pipeline depth
(7 cycles) compared to the CG-OoO processor (13 cycles) allowing faster control
mis-speculation recovery.

The last source of performance improvement in CG-OoO is the Skipahead model
where \textit{limited} out-of-order instruction scheduling within each BW is
provided. This feature leverage the Head Buffer tables.
Figure~\ref{fig:skipahead} shows the performance gain obtained via varying the
number of HB entries.  Without Skipahead, 17\% of the gap between OoO and InO is
closed.  \textit{Skipahead 2} refers to a HB with two entries; Skipahead 2
closes an additional 67\% of the performance gap between InO and OoO. Skipahead
4 (i.e. 4-entry HB) closes the rest of the performance gap. No significant
performance difference is observed for larger HB sizes. All CG-OoO results use
the statically list scheduled code.

\label{sec:res_skipahead}

\begin{figure}
	\centering
	\includegraphics[width=1.0\columnwidth]{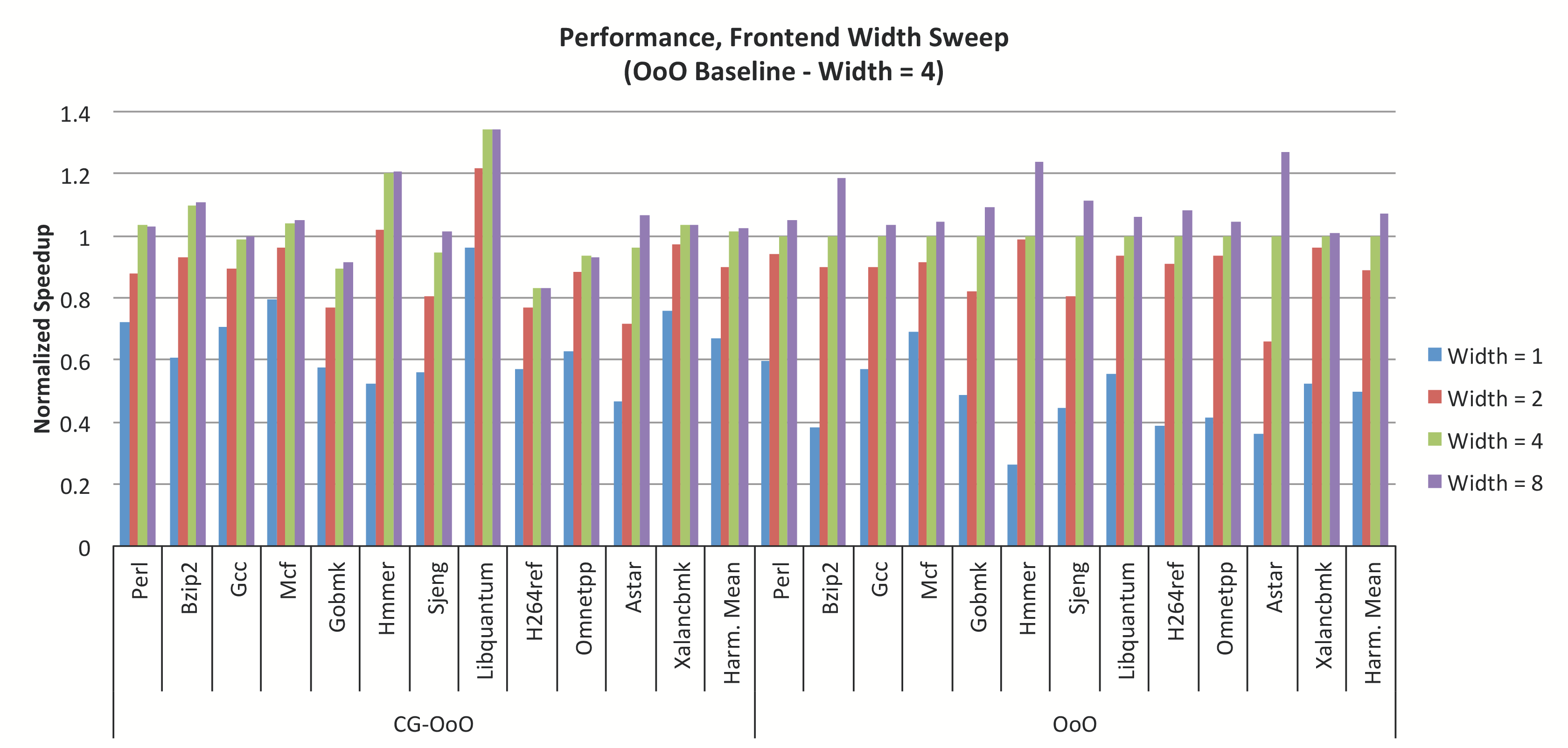}
    \caption{The speedup of the OoO and CG-OoO for different front-end widths.}
	\label{fig:perf_frontend_width}
\end{figure}

Figure~\ref{fig:perf_frontend_width} shows the CG-OoO performance as the
processor front-end width varies from 1 to 8. Comparing the harmonic mean
results for the OoO and CG-OoO shows the CG-OoO processor is superior on
narrower designs. A wider front-end delivers more dynamic operations to the
back-end. Because the OoO model has access to all in-flight operations, it can
exploit a larger effective instruction window.  Despite the larger number of
in-flight operations, the CG-OoO model maintains a limited view to the in-flight
operations making an 8-wide CG-OoO machine not much superior to its 4-wide
counterpart.

\subsection{CG-OoO Energy Analysis}
\begin{figure}[h]
\centering
    \begin{subfigure}[b]{\columnwidth}
        \includegraphics[width=\columnwidth]{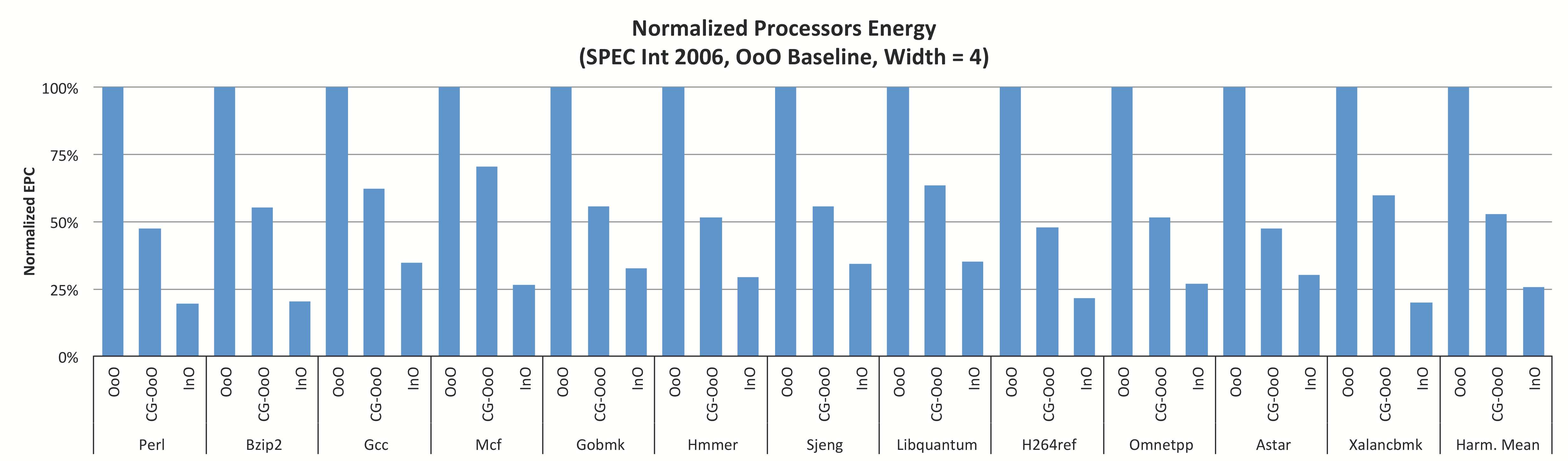}
        \caption{Normalized EPC of processors (cache energy included).}
        \label{fig:energy}
    \end{subfigure}

    \begin{subfigure}[b]{\columnwidth}
        \includegraphics[width=\columnwidth]{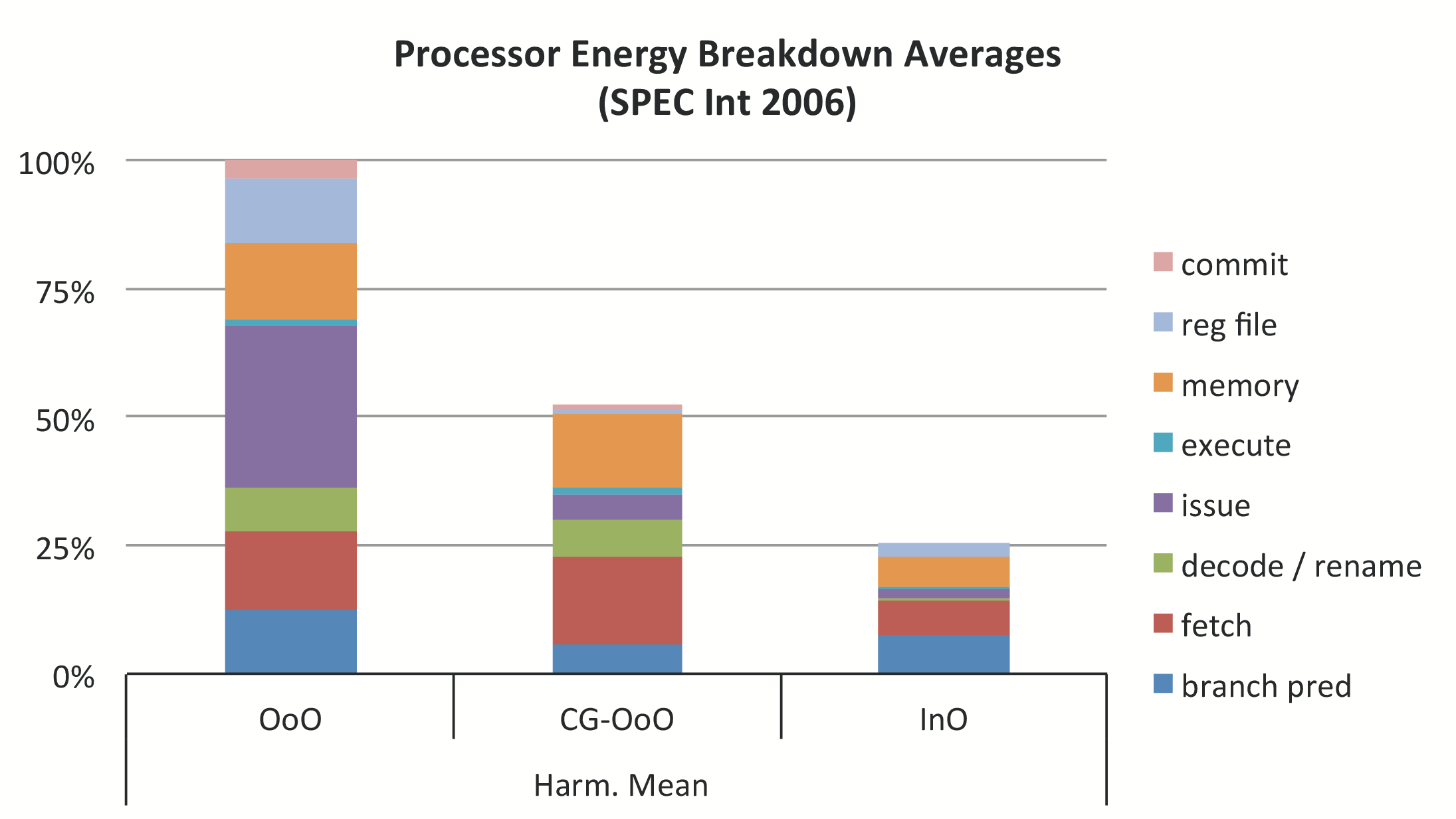} 
        \caption{Harmonic mean EPC breakdown of all processors.}
        \label{fig:energy_breakdown_mean}
    \end{subfigure}

    \begin{subfigure}[b]{\columnwidth}
        \includegraphics[width=\columnwidth]{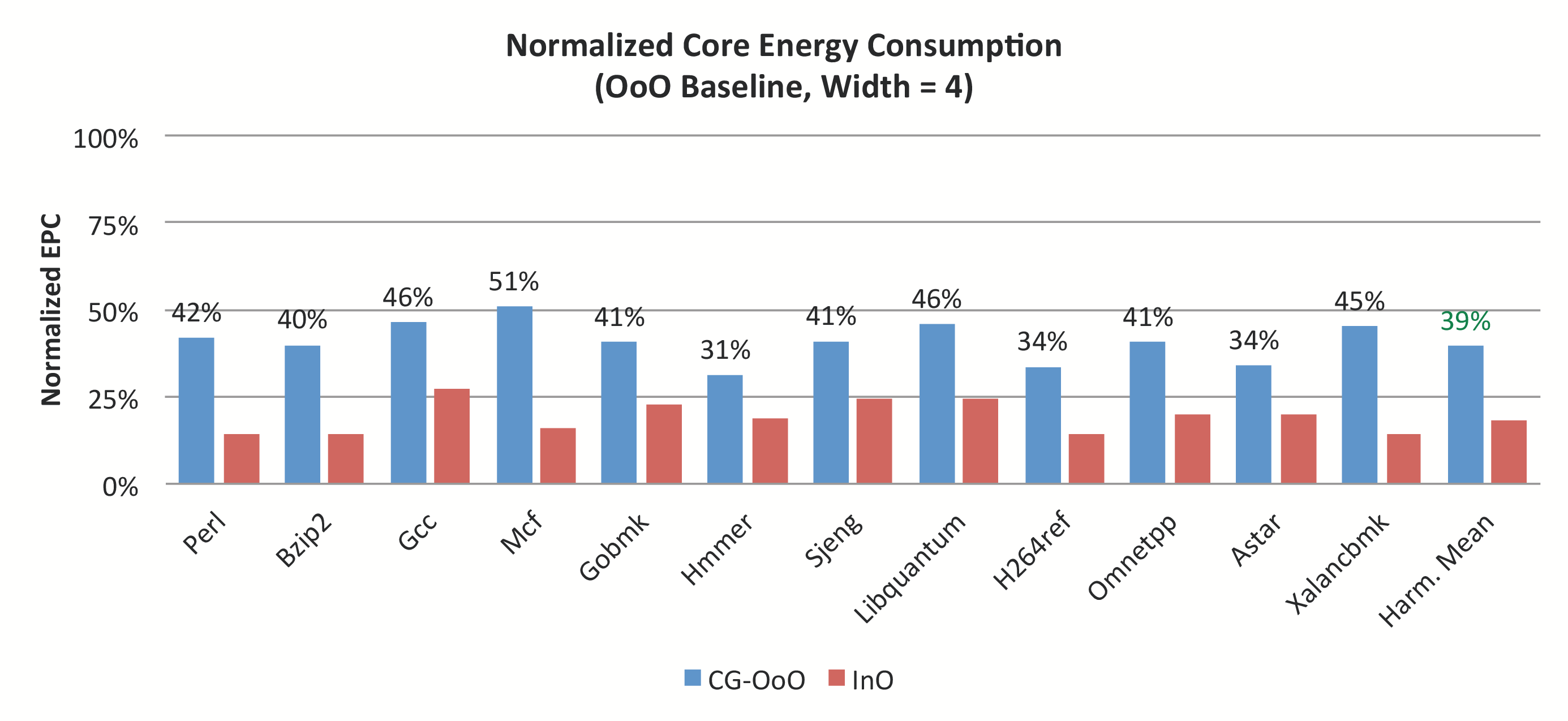} 
        \caption{Normalized EPC of processors (cache energy excluded).}
        \label{fig:energy_core}
    \end{subfigure}

    \caption{CG-OoO, OoO, InO energy per cycle (EPC).}
\end{figure}

In this section, the source of energy saving within each stage is discussed.
Overall, CG-OoO shows an average 48\% energy reduction across all benchmarks.
Energy results are measured in terms of energy per cycle (EPC).
Figure~\ref{fig:energy} shows the total energy level for the CG-OoO, OoO, and
InO processors; Figure~\ref{fig:energy_breakdown_mean} shows the harmonic mean
energy breakdown for different pipeline stages; all benchmarks follow a similar
energy breakdown trend as the harmonic mean. This figure shows the main energy
savings are in the \textit{Branch Prediction}, \textit{Register Rename},
\textit{Issue}, \textit{Register File} access, and \textit{Commit} stages.
Figure~\ref{fig:energy_core} shows 61\% average energy saving for the CG-OoO
compared to OoO baseline with similar performance. Since the main contribution
of this paper is an energy efficient processor \textit{core},
Figure~\ref{fig:energy_core} excludes cache and memory energy.

\begin{figure}[h]
	\centering
	\includegraphics[width=1.0\columnwidth]{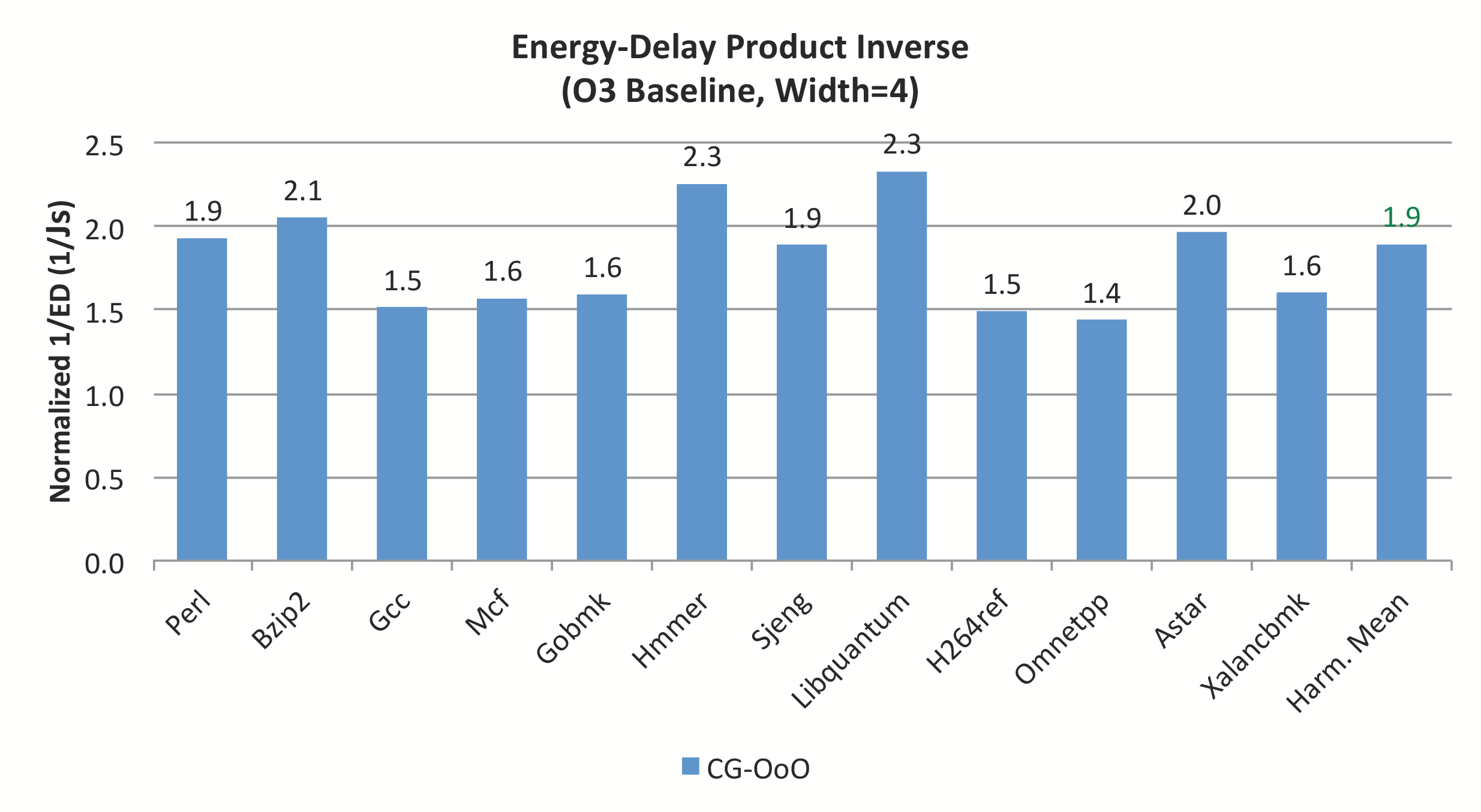}
    \caption{The CG-OoO inverse of energy-delay product normalized to OoO.}
	\label{fig:ed}
\end{figure}

Figure~\ref{fig:ed} shows the inverse of energy-delay (ED) product indicating
the favorable energy-delay characteristics of the CG-OoO over OoO for all
benchmarks, even those that fall short of the OoO performance such as Sjeng and
Gobmk. The CG-OoO is $1.9\times$ more efficient than the OoO on average.

\begin{figure}
	\centering
    \vspace {-10 pt}
	\includegraphics[width=\columnwidth]{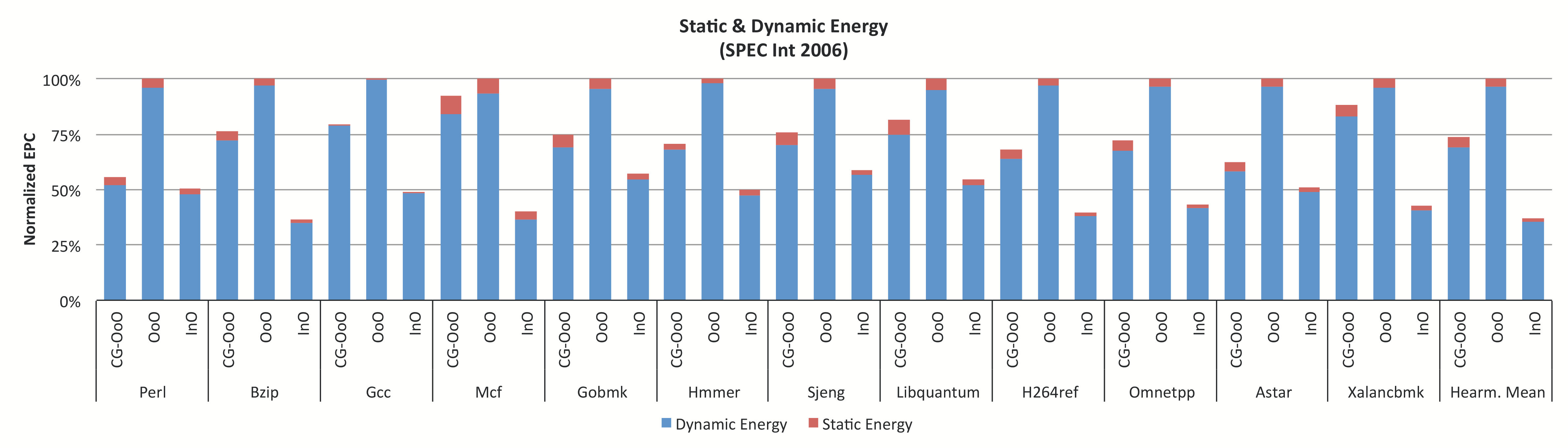}
    \caption{Static and dynamic EPC normalized to OoO.}
	\label{fig:stat_dyn_en}
\end{figure}

Figure~\ref{fig:stat_dyn_en} shows the static and dynamic energy breakdown for
different benchmarks relative to the OoO baseline. On average, the leakage
energy is smaller than 4\% of the total energy.

\subsubsection{Block Level Branch Prediction}
\begin{figure}[h]
	\centering
	\includegraphics[width=0.8\columnwidth]{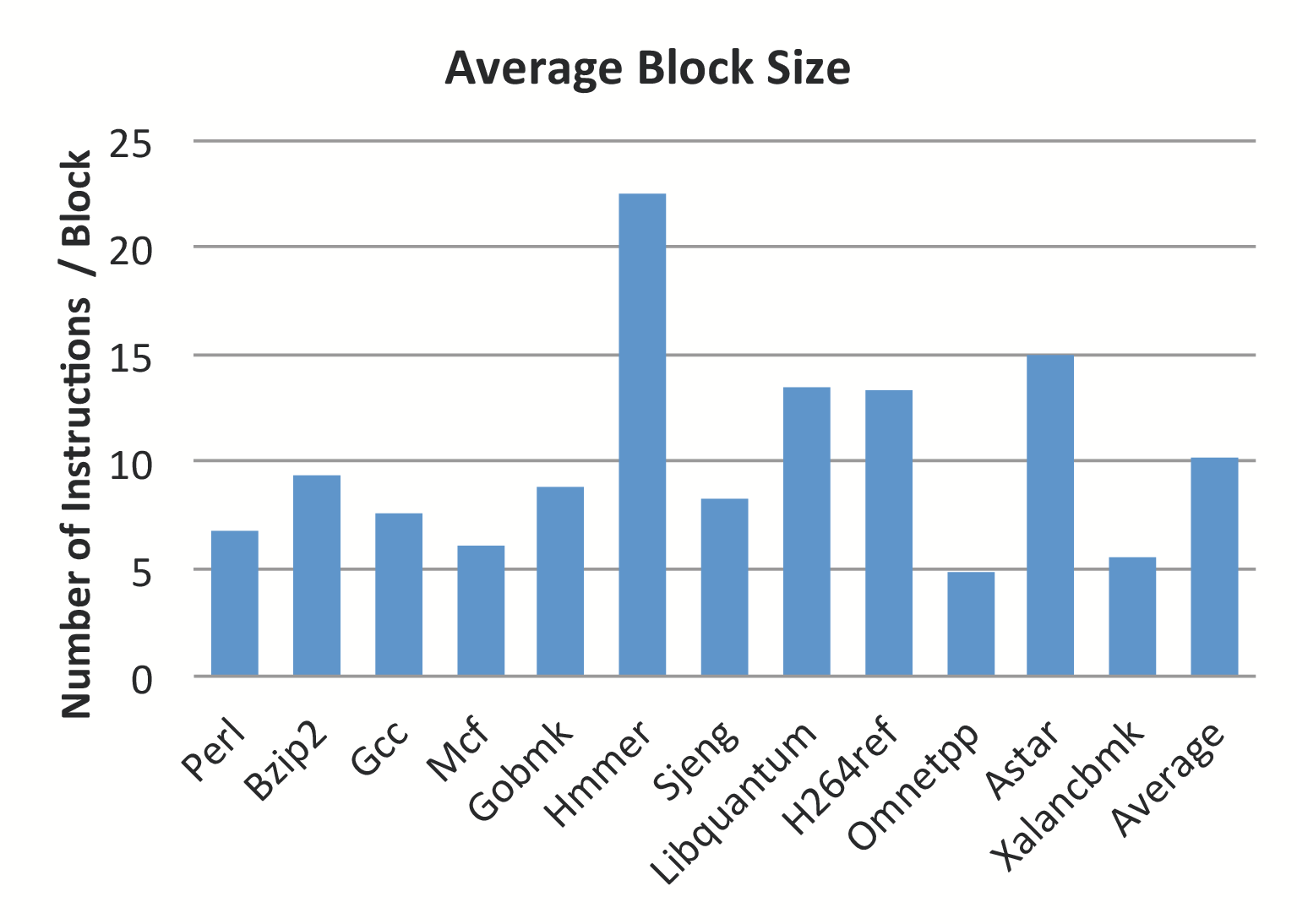} 
    \caption{Average code block size of SPEC Int 2006 benchmarks.}
	\label{fig:blk_size}
    \vspace {-10 pt}
\end{figure}

\begin{figure}[h]
	\centering
	\includegraphics[width=\columnwidth]{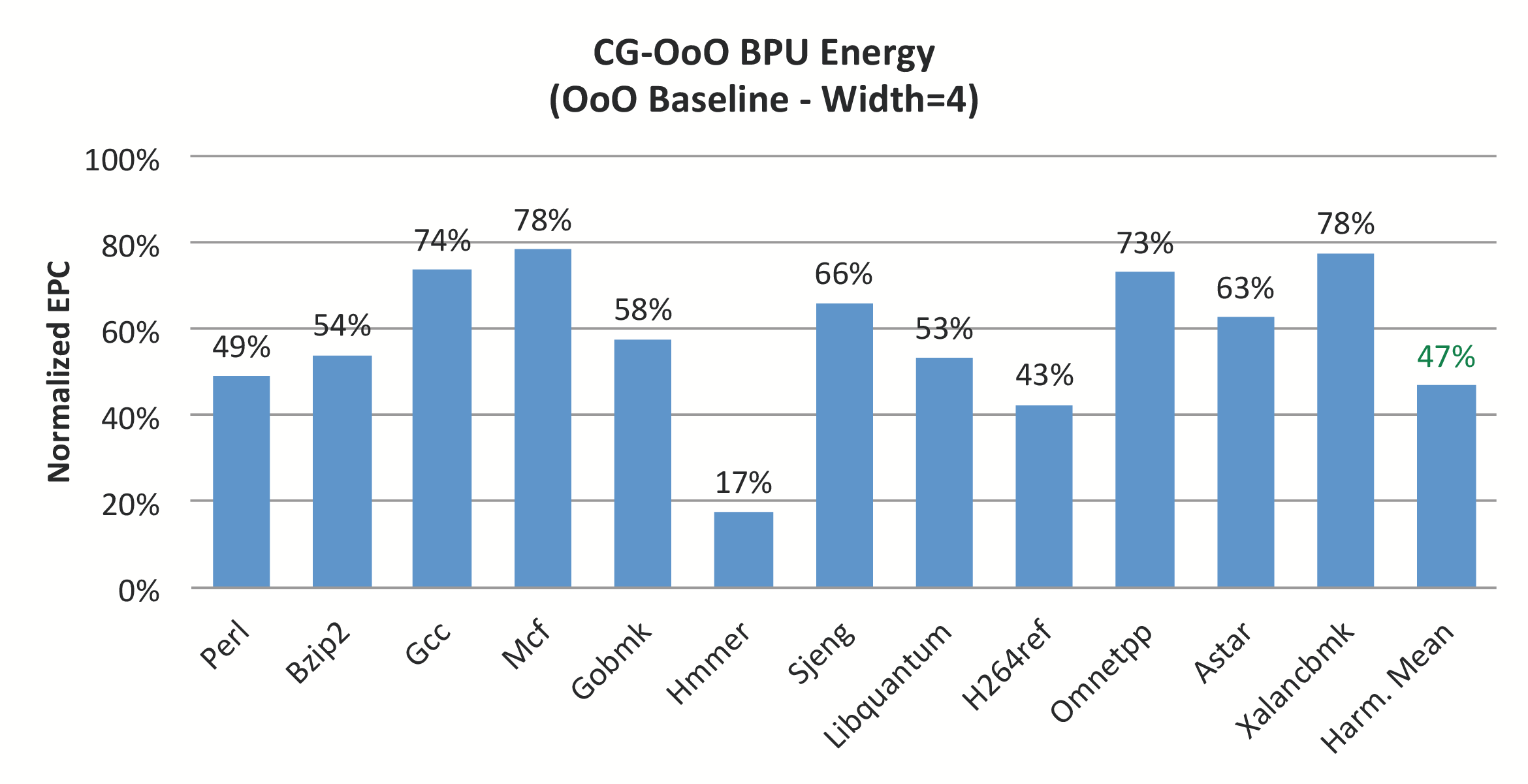} 
    \caption{The BPU access EPC normalized to OoO.}
	\label{fig:bpu}
\end{figure}

Block-level branch prediction is primarily focused on saving energy by accessing
the branch prediction unit at block granularity rather than fetch-group
granularity. Figure~\ref{fig:blk_size} shows the average block sizes for SPEC
Int 2006 benchmarks. For a benchmark application with average block size of
eight running on a 4-wide processor, this translates to roughly $2\times$
reduction in the number of accesses to the BPU tables. Figure~\ref{fig:bpu}
shows the relative energy-per-cycle for the CG-OoO model compared to the OoO
baseline.  On average, Block Level BP is 53\% more energy efficient than the OoO
model.  Hmmer shows 83\% reduction in branch prediction energy because of its
larger average code block size.

\subsubsection{Register File Hierarchy}\label{sec:reg_file_hier_energy} 
The CG-OoO register file hierarchy contributes to the processor energy savings
in four different ways, each of which is discussed here.
\begin{enumerate*}[label=(\alph*)]
    \item LRF's are low energy tables,
    \item segmented GRF reduce access energy,
    \item local operands bypass register renaming, and
    \item register renaming is optimized to reduce on-chip data movement.
\end{enumerate*}

\begin{figure}
	\centering
    \vspace {-10 pt}
	\includegraphics[width=\columnwidth]{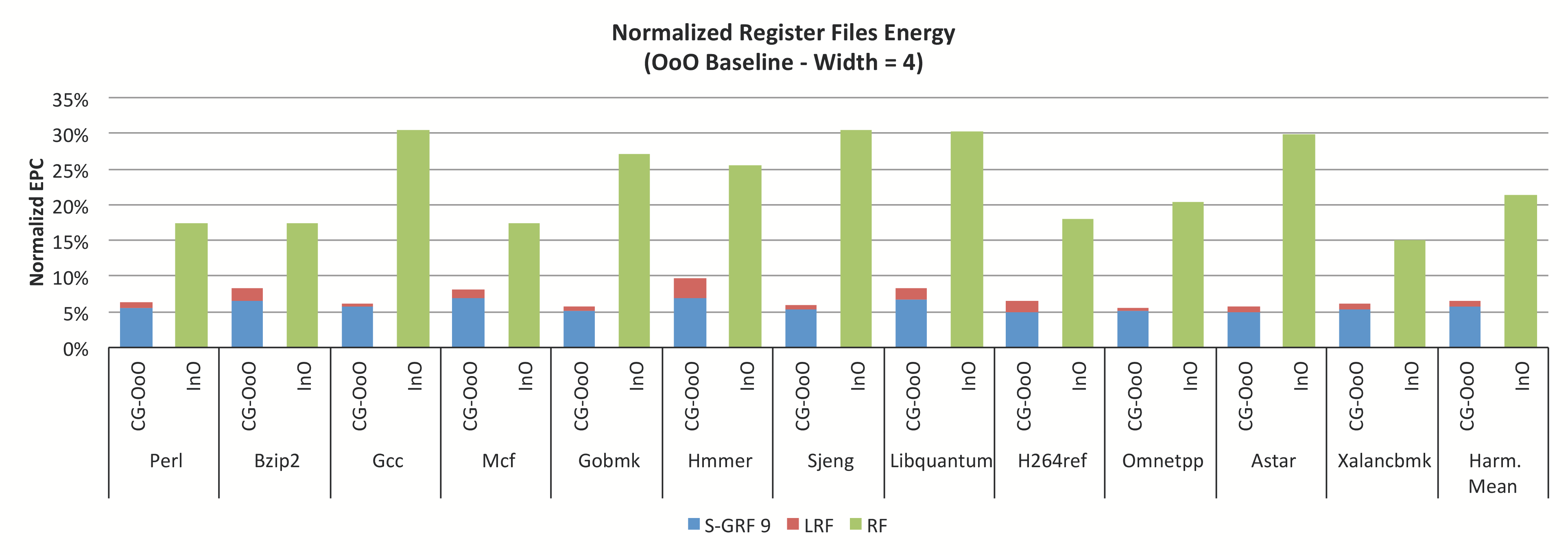} 
    \caption{The register file (RF) access EPC normalized to the OoO processor.
Overall, the CG-OoO RF hierarchy is 94\% more efficient than that of the OoO.
S-GRF 9 shows the energy of a CG-OoO GRF with 9 segments, and RF shows the
energy of an InO processor register file with the same number of ports as the
OoO processor.}
	\label{fig:rf}
\end{figure}

Local registers are statically managed, and account for 30\% of the total data
communication. The 20-entry LRF energy-per-access is about 25$\times$ smaller
than that of a unified, 256-entry register file in the baseline OoO processor.
The LRF has 2 read and 2 write ports and the unified register file has 8 read
and 4 write ports. In addition, since each BW holds a LRF near its instruction
window and execution units, operand reads and writes take place over shorter
average wire lengths. LRF's also enable additional energy saving by avoiding
local write-operand wakeup broadcasts.  Figure~\ref{fig:rf} shows the
contribution of the local register file energy compared to the OoO baseline; it
shows an average 26\% reduction in register file energy consumption due to local
register accesses.

\begin{figure}
	\centering
	\includegraphics[width=\columnwidth]{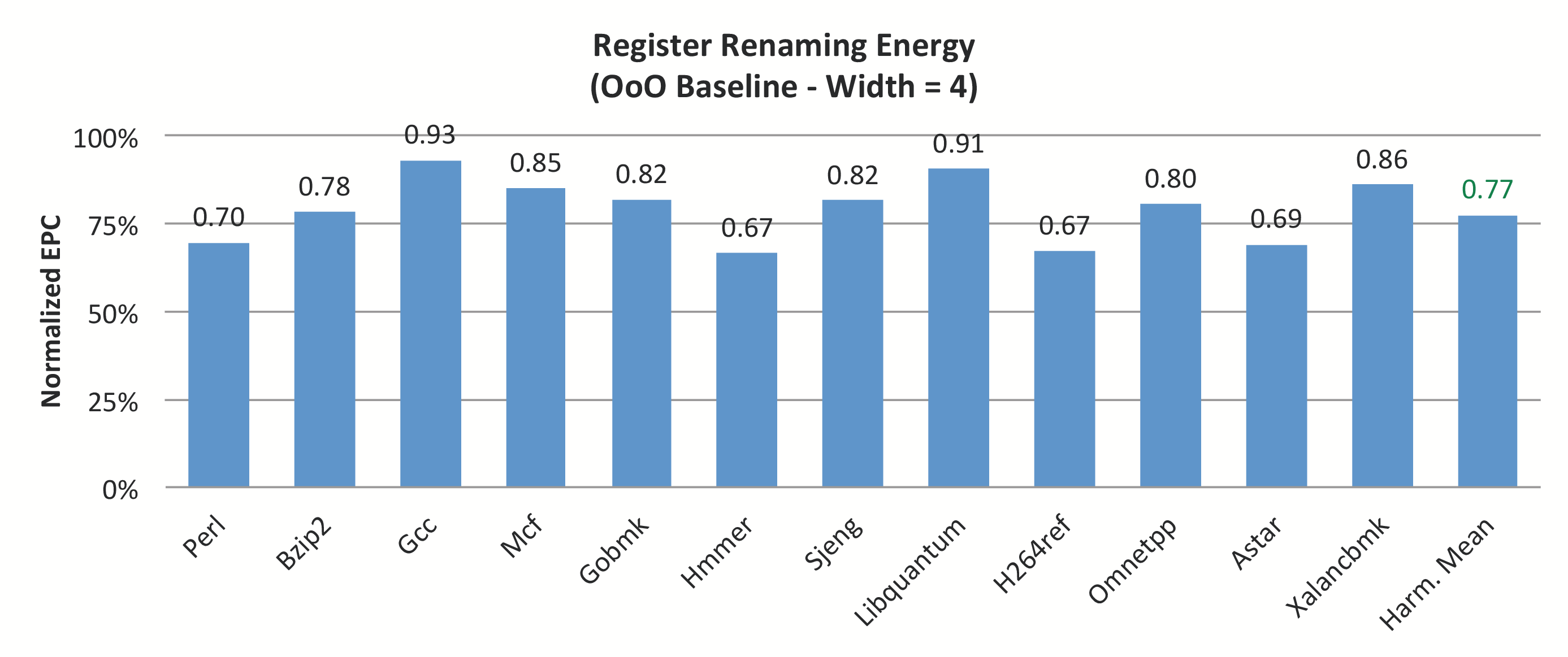} 
    \caption{The register rename access EPC normalized to OoO.}
	\label{fig:rr}
\end{figure}

Because local register operands are statically allocated, they do not require
register renaming. As a result, 23\% average energy consumption reduction is
observed in the register rename stage (see Figure~\ref{fig:rr}).

The global register file used in both OoO and CG-OoO has 256 entries. While the
use of local registers enables the use of a smaller global register file in
CG-OoO without noticeable reduction in performance, our experiments use equal
global register file sizes for fair energy and performance modeling between
CG-OoO and OoO. 

\begin{figure}
	\centering
	\includegraphics[width=\columnwidth]{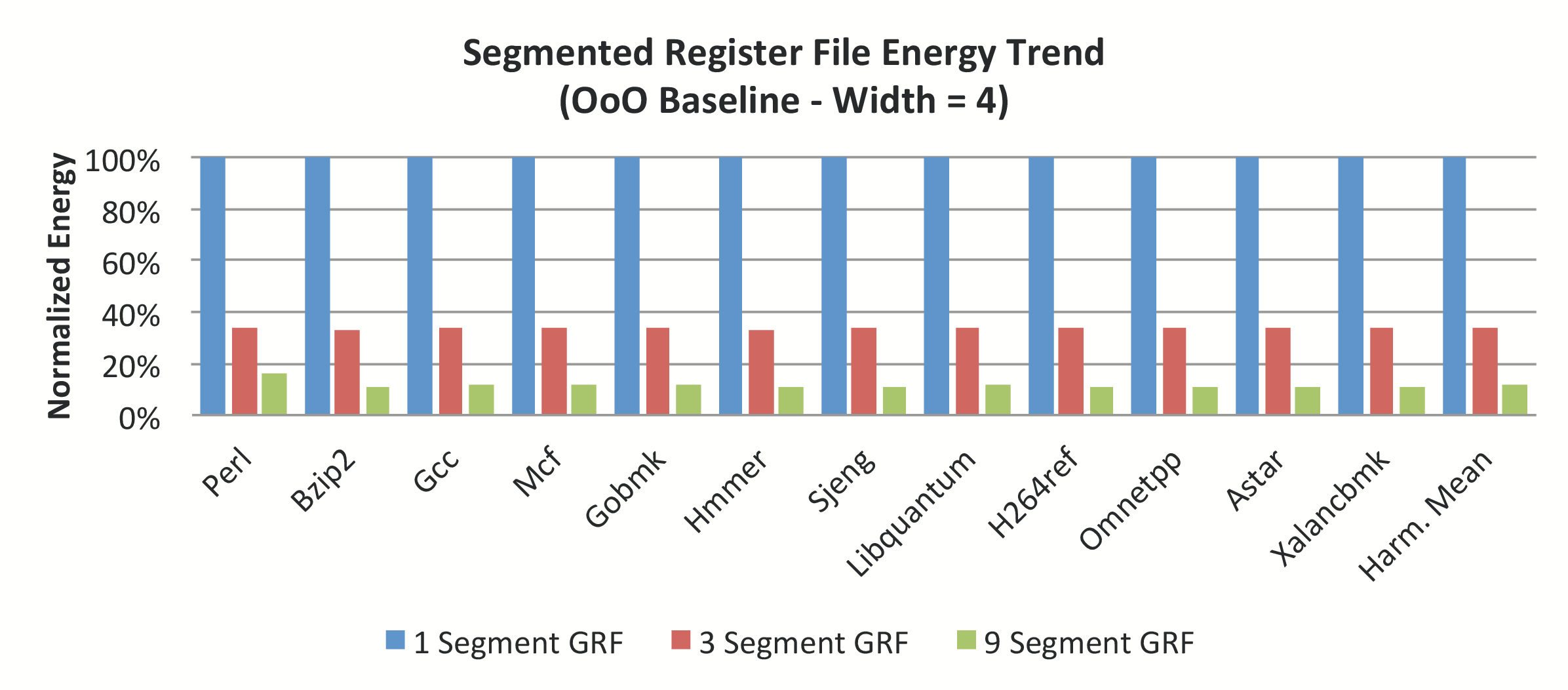} 
    \caption{The S-GRF access EPC normalized to OoO.}
	\label{fig:sgrf}
\end{figure}

To reduce the access energy overhead of a unified register file and to increase
the aggregate number of ports in the CG-OoO, this processor model breaks the
global register file (GRF) into multiple segments. Each segment is placed next
to a BW. The access energy to each register file segment is divided by the
number of segments relative to the OoO unified register file access energy.
Figure~\ref{fig:rf} also shows the contribution of the global register file
energy compared to the OoO baseline; it shows an average 68\% reduction in the
global register file energy consumption due to register file segmentation.
Notice GRF segmentation is not commonly used in OoO architectures; some ARM
architectures bank the register file for various purposes such as better thread
context switching support~\cite{arm}. Figure~\ref{fig:sgrf} shows the effect of
register file segmentation on energy.  It shows the case of a unified GRF, one
GRF segment per cluster (for a 3-cluster CG-OoO), and one GRF segment per BW. As
the number of register segments increases, energy consumption decreases
linearly.

Placing a GRF segment next to each BW is energy saving when operations
read/write global operands from/to the closest segment. Our register renaming
algorithm reduces data communication over wires by allocating an available
physical register from the GRF segment nearest to the BW of the renamed
instruction.

\subsubsection{Instruction Scheduling} 
\begin{figure}[h]
	\centering
	\includegraphics[width=\columnwidth]{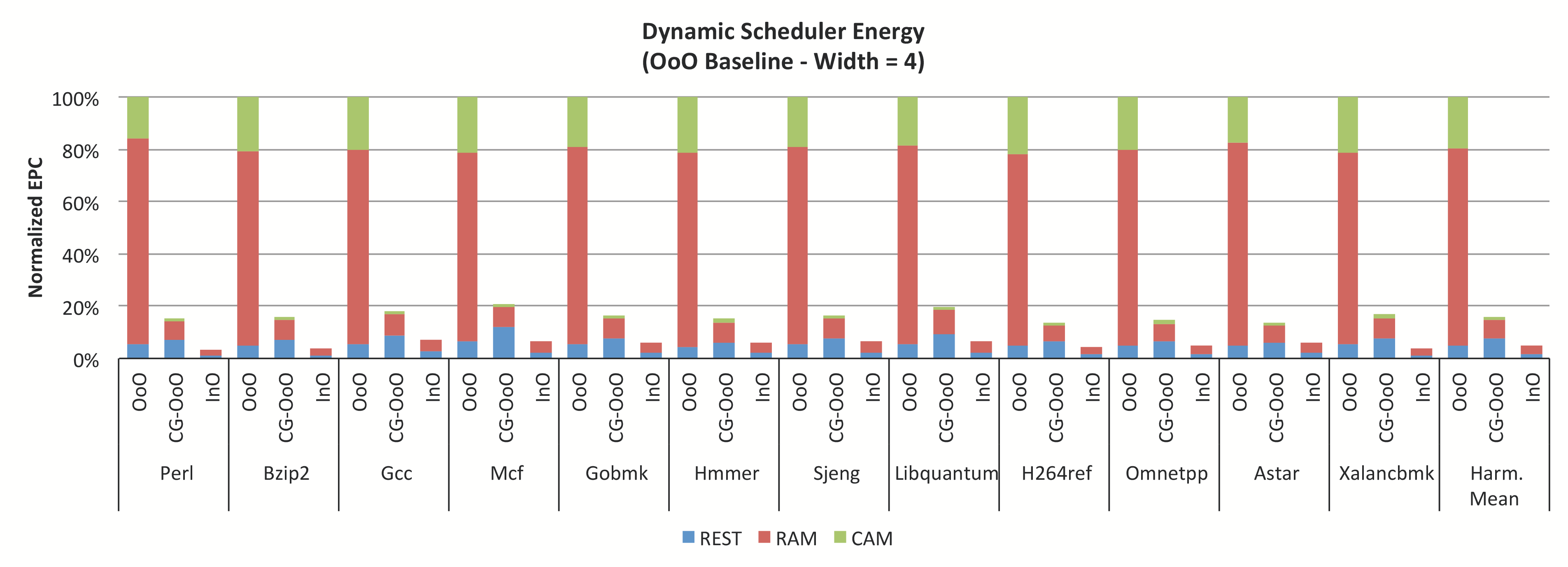} 
    \caption{The instruction issue EPC normalized to OoO.}
	\label{fig:schedule}
\end{figure}

The CG-OoO processor introduces the Skipahead issue model. In OoO and CG-OoO,
in-flight instructions are maintained in queues that are partly RAM and partly
CAM tables. For the InO model, instructions are held in a small FIFO buffer.
Figure~\ref{fig:schedule} shows the energy breakdown of the dynamic scheduling
hardware; it shows the majority of the OoO scheduling energy (75\%) is in
reading and writing instructions from the RAM table. Another 20\% of the OoO
energy is in CAM table accesses. The ``\textit{Rest}" of the energy is consumed
in stage registers and the interconnects used for instruction wakeup and select.
This figure also indicates 90\% average reduction in the CG-OoO RAM table energy
(relative to OoO RAM energy) which is due to accessing smaller SRAM tables, and
95\% average reduction in the CAM table energy which is due to using 2 to
4-entry Head Buffers (HB) instead of the 128-entry CAM tables used in the
baseline OoO instruction queue. The ``Rest" average energy is increased by 40\%
due to the more pipeline registers at the issue stage. Overall, the CG-OoO issue
stage is 84\% more efficient than OoO. 

\subsubsection{Block Re-Order Buffer}
\begin{figure}[h]
	\centering
	\includegraphics[width=\columnwidth]{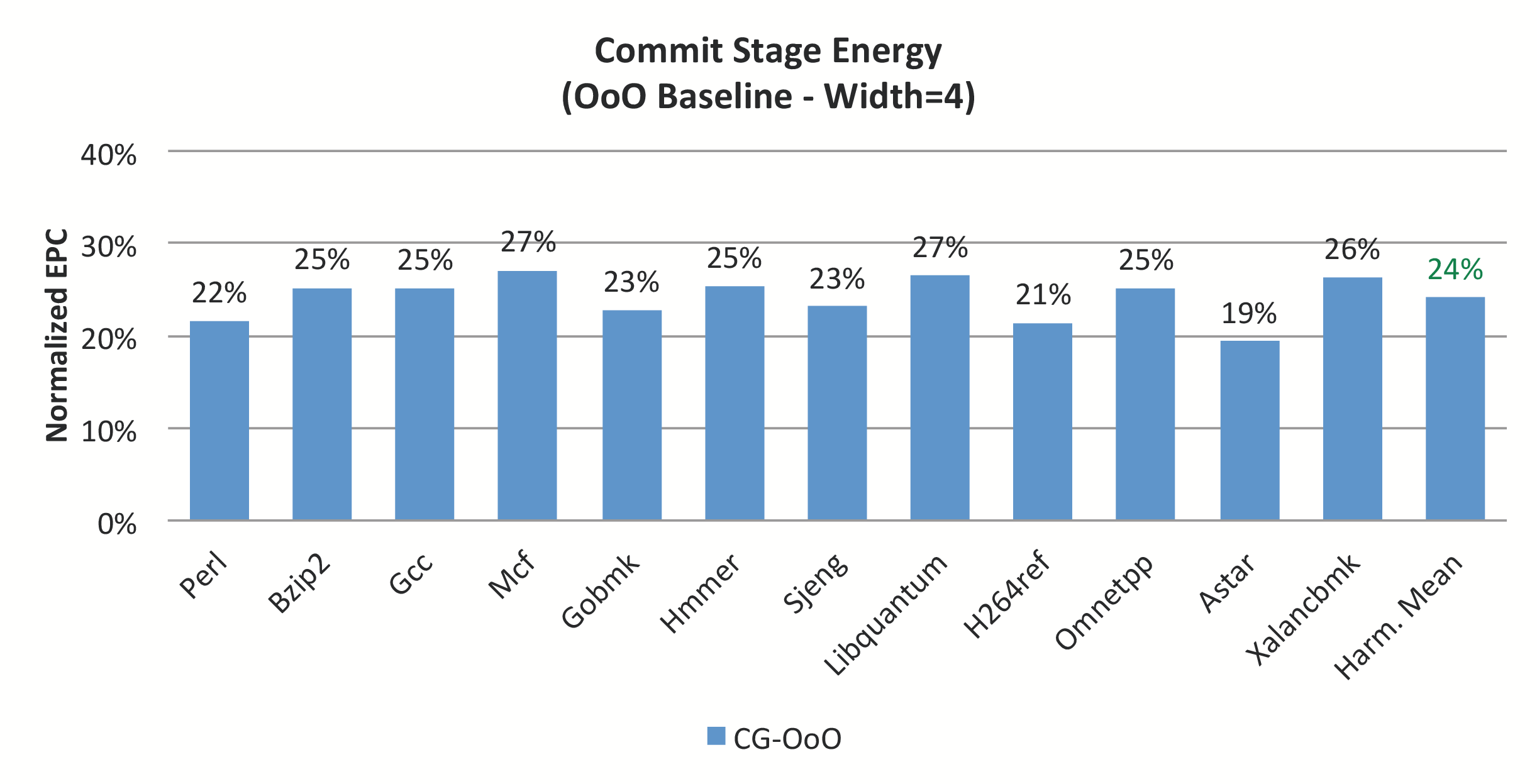} 
    \caption{The commit EPC normalized to OoO.}
	\label{fig:commit}
\end{figure}

The CG-OoO processor maintains program order at block-level granularity. This
makes read-write accesses to the BROB substantially smaller than that of OoO
ROB. Block write operations are done after decoding each \texttt{head} and block
reads are done at the commit stage. Instructions access BROB to notify the
corresponding block entry of their completion. In addition, since the BROB is
designed to maintain program order at block granularity, it is provisioned to
have 16 entries rather than 160 entries used for
OoO~\cite{coorporation2009intel}. The $10\times$ reduction in the re-order
buffer size makes all read-write operations $10\times$ less energy consuming.
Figure~\ref{fig:commit} shows 76\% average energy saving for CG-OoO.

\subsection{Clustering and Scaling Analysis}
The CG-OoO architecture focuses on reducing processor energy through designing a
complexity-effective architecture; to remain competitive with the OoO
performance, this architecture supports a larger number of execution units (EU).
To do so, the CG-OoO model must employ a design strategy that is more scalable
than the OoO. A cluster consists of a number of BW's sharing a number of EU's.
To illustrate the effect of different clustering configurations, the
experimental results in this section assume three clusters.  

\begin{figure}
\centering
    \begin{subfigure}[b]{0.48\columnwidth}
        \includegraphics[width=\columnwidth]{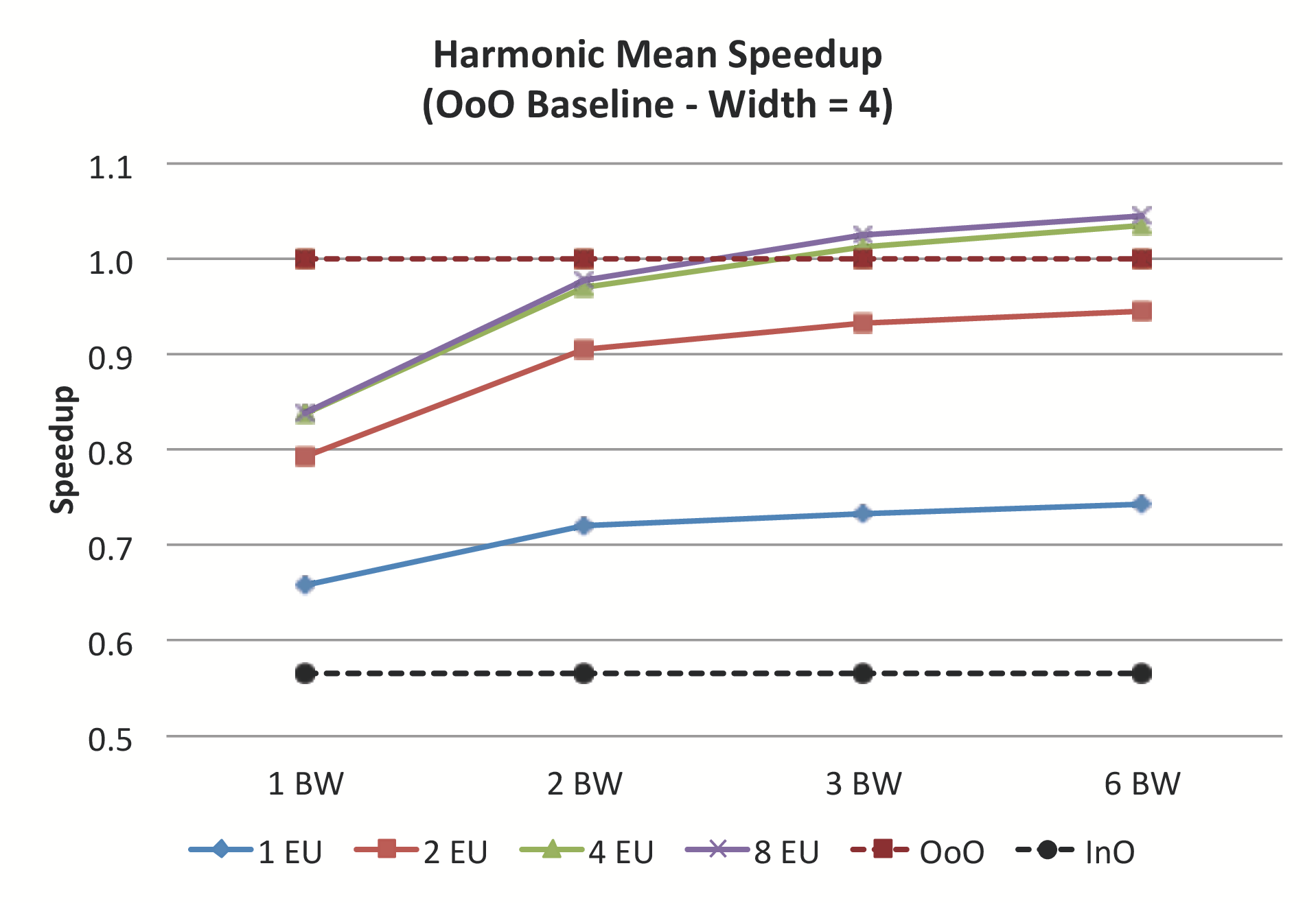} 
        \caption{}
        \label{fig:cluster_perf}
    \end{subfigure}
    ~
    \begin{subfigure}[b]{0.48\columnwidth}
        \includegraphics[width=\columnwidth]{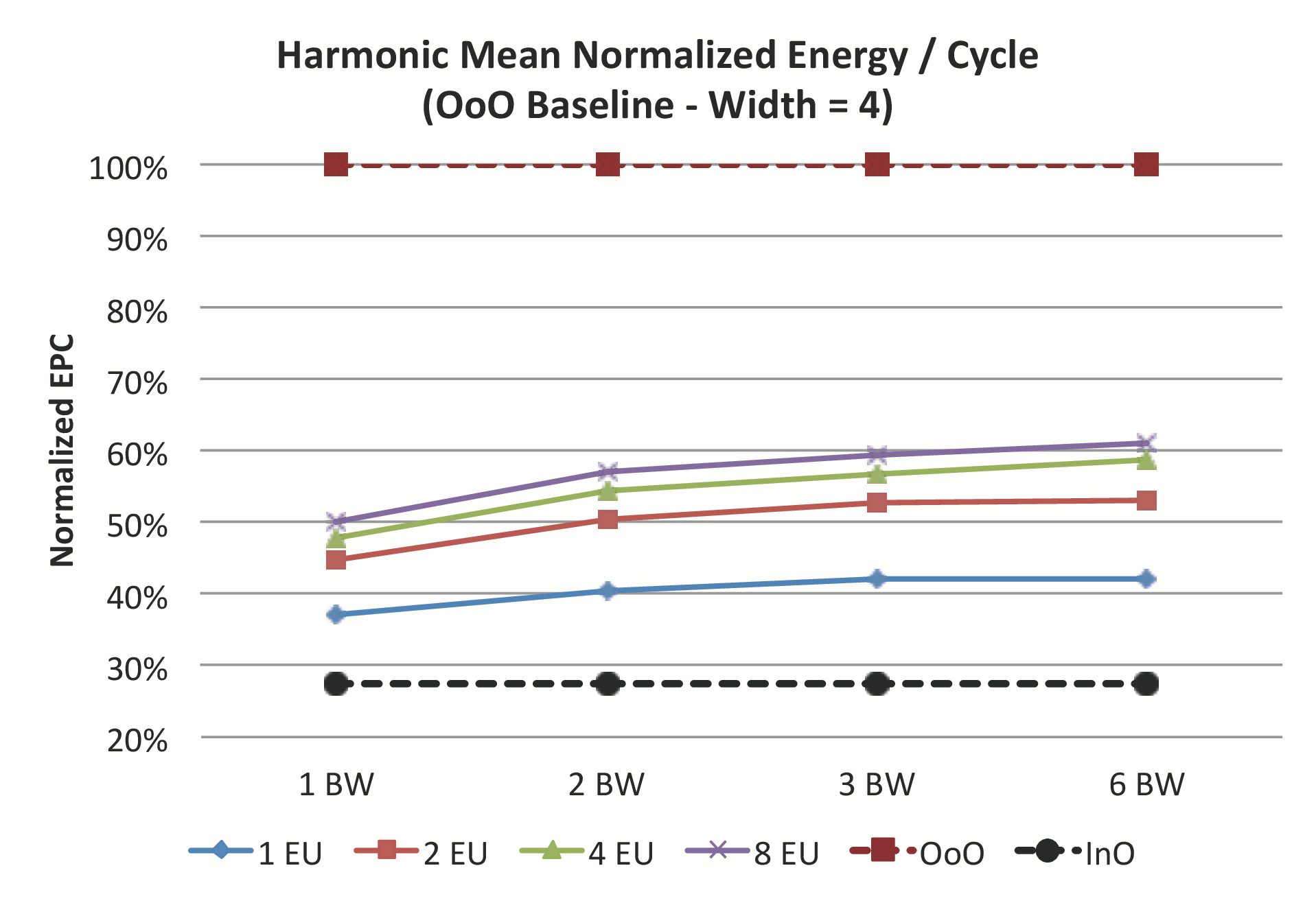} 
        \caption{}
        \label{fig:cluster_enrg}
    \end{subfigure}
    \caption{Normalized Performance \& Energy for different clustering
configurations. All configurations assume a 3-cluster CG-OoO model; the total
number of BW's and EU's is calculated through multiplying the above numbers by
3. Here, performance is measured as the harmonic mean of the IPC and the energy
is measured as the harmonic mean of the EPC over all the SPEC Int 2006
benchmarks.}
\end{figure}

Figures~\ref{fig:cluster_perf} and \ref{fig:cluster_enrg} show the normalized
average performance and energy of SPEC Int 2006 benchmarks versus the number of
BW's per cluster for various number of EU's per cluster. The speedup figure
shows some clustering configurations reach beyond the performance of the OoO.
All clustering models exhibit substantially lower energy consumption overhead
compared to the OoO design. The most energy efficient configuration is the one
with 1 BW and 1 EU per cluster; it is 63\% more energy efficient than the OoO,
but only at 65\% of the OoO performance. The most high-performance configuration
evaluated here is the one with 6 BW's and 8 EU's per cluster; it is 39\% more
energy efficient than the OoO, and 104\% of the OoO performance. The design
configuration evaluated throughout this section corresponds to the cross-over
performance point with 3 BW's and 4 EU's per cluster.

\begin{figure}
	\centering
    \vspace {-10 pt}
	\includegraphics[width=\columnwidth]{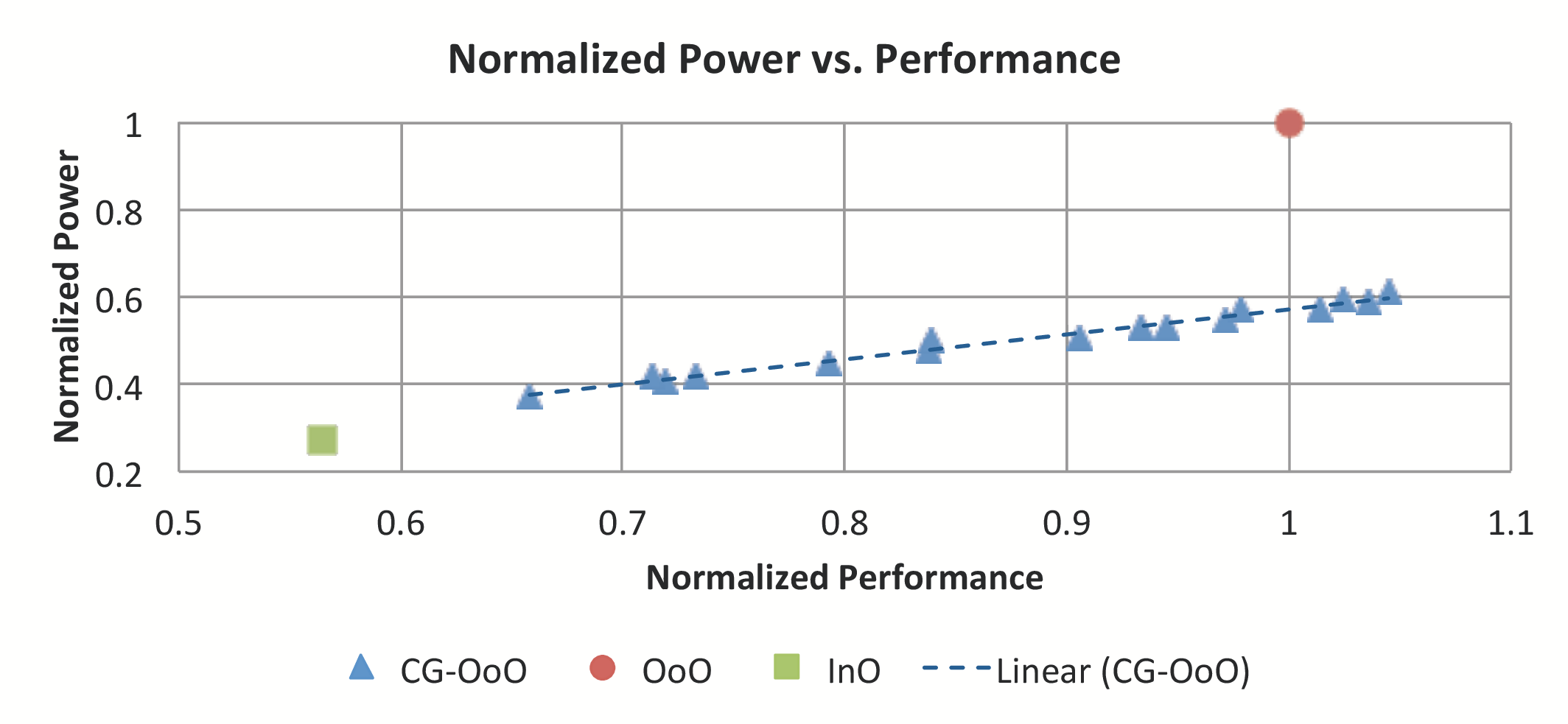}
    \caption{The energy versus performance plot showing different CG-OoO
configurations normalized to the OoO. The CG-OoO core configurations illustrate
the energy-performance proportionality attribute of the CG-OoO. Performance is
measured as the harmonic mean of the IPC and energy is measured as the harmonic
mean of the EPC over the SPEC Int 2006 benchmarks.}
	\label{fig:ep_trend}
\end{figure}

Figure~\ref{fig:ep_trend} shows the energy-performance characteristics of the
CG-OoO model plotting all the cluster configurations presented above.  The
lowest energy-performance point in the plot refers to the 1 BW, 1EU per cluster
configuration and the highest energy-performance point refers to the 6 BW, 8 EU
per cluster configuration. This figure suggests as the processor resource
complexity increases, the energy-performance characteristics grow
proportionally. Beyond a certain scaling point, the wakeup/select and load-store
unit wire latencies become so large that the energy-performance proportionality
of the CG-OoO will break. Identifying this energy-performance point is outside
of the scope of this work. 

\section{Conclusion}\label{sec:conclusion}
The CG-OoO leverages a distributed micro-architecture capable of issuing
instructions from multiple code blocks concurrently. The key enablers of energy
efficiency in the CG-OoO are (a) its end-to-end complexity effective design, and
(b) its effective use of compiler assistance in doing code clustering and
generating efficient static code schedules.  Despite the reliance of the CG-OoO
architecture in providing energy efficiency static code, it requires no
profiling. This architecture is an energy-proportional design capable of scaling
its hardware resources to larger or smaller computation units according to the
workload demands of programs at runtime. The CG-OoO supports an out-of-order
issue model at block granularity and a limited out-of-order issue model at
instruction granularity (i.e. within block).  It leverages a hierarchical
register file model designed for energy efficient data transfer. Unlike most
previous studies, this work performs a detailed processor energy modeling
analysis. CG-OoO reaches the performance of the out-of-order execution model
with over 50\% energy saving.



\clearpage
\bibliographystyle{ieeetr}
\bibliography{references}

\end{document}